\documentclass[10pt,preprint]{aastex}
\usepackage{multicol}
\usepackage{color}
\usepackage{soul}
\usepackage[normalem]{ulem}

\newcommand{\be}{\begin{eqnarray}} 
\newcommand{\ee}{\end{eqnarray}} 
\newcommand{\Msun}{\mbox{$M_{\odot}\;$}}

\def\simge{\mathrel{\rlap{\raise 0.511ex
     \hbox{$>$}}{\lower 0.511ex \hbox{$\sim$}}}}
\def\simle{\mathrel{\rlap{\raise 0.511ex
      \hbox{$<$}}{\lower 0.511ex \hbox{$\sim$}}}}

\begin{document}

\title{Neutron Star Masses and Radii from Quiescent Low-Mass X-ray
  Binaries}

\author{James M. Lattimer}
\affil{Department of Physics and Astronomy, State University of New York 
at Stony Brook, Stony Brook, NY 11794-3800, USA}
\email{james.lattimer@stonybrook.edu}

\author{Andrew W. Steiner}
\affil{Institute for Nuclear Theory, University of Washington,
Seattle, WA 98195, USA}
\email{steiner3@uw.edu}

\shorttitle{Neutron Star Masses and Radii}
\shortauthors{LATTIMER \& STEINER}

\begin{abstract}
  We perform a systematic analysis of neutron star radius constraints
  from five quiescent low-mass X-ray binaries and examine how they
  depend on measurements of their distances and amounts of intervening
  absorbing material, as well as their assumed atmospheric
  compositions. We construct and calibrate to published results a
  semi-analytic model of the neutron star atmosphere which
  approximates these effects for the predicted masses and radii.
  Starting from mass and radius probability distributions established
  from hydrogen-atmosphere spectral fits of quiescent sources, we
  apply this model to compute alternate sets of probability
  distributions. We perform Bayesian analyses to estimate neutron star
  mass-radius curves and equation of state (EOS) parameters that
  best-fit each set of distributions, assuming the existence of a
  known low-density neutron star crustal EOS, a simple model for the
  high-density EOS, causality, and the observation that the neutron
  star maximum mass exceeds $2~M_\odot$. We compute the posterior
  probabilities for each set of distance measurements and assumptions
  about absorption and composition. We find that, within the context
  of our assumptions and our parameterized EOS models, some absorption
  models are disfavored. We find that neutron stars composed of
  hadrons are favored relative to those with exotic matter with strong
  phase transitions. In addition, models in which all five stars have
  hydrogen atmospheres are found to be weakly disfavored. Our most
  likely models predict neutron star radii that are consistent with
  current experimental results concerning the nature of the
  nucleon-nucleon interaction near the nuclear saturation density.
\end{abstract}


\section{INTRODUCTION}

Although the masses of at least 3 dozen neutron stars have been
relatively precisely measured (see \cite{Lattimer12} for a summary),
estimates of individual neutron star radii are poorly known.
Additionally, simultaneous mass and radius measurements for the same
object are relatively uncertain. The two leading candidates for such
measurements are bursting neutron stars that show photospheric radius
expansion (hereafter PRE; \citealt{vanParadijs79, Lewin93}) and
transiently accreting neutron stars in quiescence \citep{Rutledge99},
often referred to as quiescent low-mass X-ray binaries (QLMXBs). The
former class of sources are believed to be energetic enough to
temporarily lift the neutron star atmosphere off the surface and thus
have peak luminosities near the Eddington Limit, $L_{\rm Edd}=4\pi
cGM\kappa^{-1}$ where $M$ is the neutron star mass and $\kappa$ is the
atmospheric opacity. The luminosity on the tail of the burst, on the
other hand, is due to thermal radiation from the cooling star, $L=4\pi
f_c^{-4}R^2T_{\mathrm{eff}}^4$, where $R$ is the neutron star radius,
$T_{\mathrm{eff}}$ is the effective blackbody temperature, and $f_c$
is a color-correction factor representing the fact that the atmosphere
is not a true blackbody. Observations measure fluxes, so measurements
of the luminosities require knowledge of the source distances $D$. In
addition, both the Eddington Limit and the observed luminosities must
be corrected for gravitational redshift. These sources thus have the
potential to measure two combinations of mass and radius.

The currently accepted theoretical explanation for the thermal X-ray
emission from many QLMXBs is thermal energy deposited in the deep
crust during outbursts \citep{Brown98}.
QLMXBs also exhibit hard nonthermal components (e.g.,
\citealt{Campana98}), and the role played by low-level accretion is
not well-understood (e.g., \citealt{Rutledge02a,Fridriksson10}). We
ignore these potential complications in the current work and assume
that this thermal energy further diffuses into and heats the core.
Following outbursts, the energy is re-radiated on core-cooling
timescales through the neutron star's atmosphere. Since accretion of
heavier elements leads to their gravitational settling on timescales
of seconds \citep{Romani87,Bildsten92}, the atmosphere is usually
assumed to be pure hydrogen. Pure H atmosphere models
\citep{Rajagopal96, Zavlin96, McClintock04, Heinke06, Haakonsen12}
applied to QLMXBs often predict neutron star radii compatible with
theoretical radius estimates, compared, for example, to blackbody
models which would suggest $R\simle1$ km \citep{Rutledge99}.
Atmospheres dominated by heavier elements could occur when no hydrogen
is accreted, such as is the case in some ultracompact X-ray binary
systems like 4U 1820$-$30 where the donor is a helium white
dwarf~\citep{Stella87,Rappaport87}.  Recent work on helium
atmospheres by \citep{Servillat12} and \citet{Catuneanu13} based on
models from \citet{Ho09} have begun to clarify how they impact radius
constraints in QLMXBs. QLMXBs are observed both in our Galaxy and in
globular clusters, but distance uncertainties (of order 50\%)
associated with field sources precludes using them for accurate radius
estimates. As opposed to PRE sources, QLMXBs primarily
allow one to measure the angular diameter
$R_\infty/D$, where $R_\infty=R/\sqrt{1-2GM/Rc^2}$ and, to much lesser
accuracy through its weaker dependence on observed spectra, the
gravitational redshift $z=R_\infty/R-1$. Below, we always use $R$
to refer to the physical radius, not the radiation radius, which
will always be written as $R_{\infty}$.

Both types of measurements have large systematic errors, including
their distances, the amount of interstellar absorption, and their
atmospheric compositions. In addition, PRE burst sources have been
largely modeled
\citep{Ozel09,Guver10a,Guver10b,Ozel10,Steiner10,Ozel11,Suleimanov11,Steiner13,Guver13}
with static, spherically symmetric atmospheres although they are
certainly non-spherical, dynamical events. Another advantage of QLMXBs
is that their atmospheric composition is likely to be pure H (or He,
in the case of ultra-compact binaries) due to the rapid gravitational
settling time of heavier elements in their atmospheres, while PRE
atmospheres could be dominantly H or He and also have metallicities up
to solar proportions.

A recent study of five QLMXBs by \cite{Guillot13}, hereafter G13,
found that the best-fit physical radii of these five sources
ranged from 6.6 km to 19.6 km if
modeled with H atmospheres (see Table 4 in G13). Under the
assumption that all the neutron stars have hydrogen atmospheres and
the same radius, G13 determined this radius to be
$R=9.1^{+1.3}_{-1.5}$ km to 90\% confidence. Given the importance of
well-measured neutron star radii for both astrophysics and nuclear
physics~\citep{Lattimer01, Steiner12}, we reconsider these five QLMXBs
in the present study.
\section{DATA AND FITS FOR QLMXBs}
\label{Sec:Data}

In this study, we will focus on the five QLMXB sources studied by G13,
namely, the sources in M28, NGC 6397, M13, $\omega$ Cen, and NGC 6304.
The X-ray spectra of at least two other sources, X7 in 47 Tuc
\citep{Heinke06} and NGC 6553 \citep{Guillot11b}, have been previously
studied to generate mass and radius constraints. However, observations
of the neutron star X7 in 47 Tuc are affected by pileup and those of
the neutron star in NGC 6553 are contaminated from a nearby source
(see discussion in G13). S. Guillot kindly provided us the $(R,
M)$ probability distributions from the G13 spectral fits (these
distributions did not include the distance uncertainty and did not
assume that all neutron stars have the same radius).

\begin{deluxetable}{clll}
\tablecaption{Distance Measurements of QLMXBs\label{tab:dist}}
\tablehead{
\colhead{Source} &
\colhead{$D$ (kpc)} &
\colhead{$D$ (kpc)} &
\colhead{$D$ (kpc)}\\
\colhead{Label} &\colhead{ G13 } & \colhead{  Alt } & \colhead{ H10 }
}
\startdata
M28          & \phn$5.5\pm0.3$   & \phn$5.5\pm0.3$  & $5.5\pm0.3$ \\
NGC 6397     & $2.02\pm0.18$     & $2.53\pm0.1$     & $2.3\pm0.18$ \\
M13          & \phn$6.5\pm0.6$   & \phn$7.7\pm0.6$  & $7.1\pm0.6$ \\
$\omega$ Cen & \phn$4.8\pm0.3$   & \phn$4.8\pm0.3$  & $5.2\pm0.3$ \\
NGC 6304     & $6.22\pm0.26$     & $6.22\pm0.26$    & $5.9\pm0.26$ \\
\enddata
\tablecomments{Distance sets utilized in this work, with 68\% uncertainties, are labeled G13 from \cite{Guillot13}, Alt (see text), and H10 from the Harris catalog \citep{Harris10} (with the
  same uncertainties adopted by G13).}
\end{deluxetable}

\begin{deluxetable}{cllllll}
\tablecaption{Hydrogen Column Densities, Redshifts 
and Radiation Radii of QLMXBs\label{tab:nh}}
\tablehead{
\colhead{Source} &
\colhead{$N_{H21}$} &
\colhead{$R_\infty\,$(km)} &
\colhead{Ref.\hspace*{0.5cm}} &
\colhead{$N_{H21}$} &
\colhead{$R_\infty\,$(km)} &
\colhead{$z$}
}
\startdata
M28 & \,$2.5\pm0.3$ & $12.2\pm3.0$ & (1) & $2.52$ (G13) & 
$12.91_{-1.52}^{+1.47}$ & $0.199_{-0.114}^{+0.307}$ \\[2pt]
& \,$2.6\pm0.4$ & $14.5^{+6.9}_{-3.8}$ & (2) & 1.89 (D90) & 
$10.67_{-1.16}^{+1.29}$ & $0.212_{-0.116}^{+0.172}$ \\[2pt]
& & & & 2.74 (H10)$\!$ & 
$13.66_{-1.52}^{+1.63}$ & $0.226_{-0.141}^{+0.324}$ \\[2pt]
\hline\\*[-8pt]
NGC 6397 & 1.4 & $9.6^{+0.8}_{-0.6}$ & (3) & $0.96$ (G13) & 
$8.58_{-1.31}^{+1.30}$ & $0.197_{-0.067}^{+0.083}$ \\[2pt]
& & & & 1.4 (D90) & 
$11.68_{-1.62}^{+1.96}$ & $0.240_{-0.104}^{+0.215}$ \\[2pt]
& & & & 1.23 (H10) $\!$ & 
$10.55_{-1.54}^{+1.71}$ & $0.238_{-0.101}^{+0.154}$ \\[2pt]
\hline\\*[-8pt]
M13 & 0.11 & $10.8\pm0.3$ & (4) & $0.08$ (G13) & 
$11.56_{-2.16}^{+2.60}$ & $0.274_{-0.171}^{+0.183}$ \\
& 0.11 & $12.3^{+1.4}_{-1.7}$ & (5) & 0.145 (D90) & 
$13.07_{-2.51}^{+2.84}$ & $0.285_{-0.184}^{+0.235}$ \\
& 0.12 & $10.6^{+0.3}_{-0.4}$ & (6) & 0.137 (H10) $\!$ & 
$12.81_{-2.43}^{+2.79}$ & $0.283_{-0.180}^{+0.227}$ \\
\hline\\*[-8pt]
$\omega$ Cen & 0.9 & $13.7\pm2.0$ & (7) & $1.82$ (G13) & 
$23.03_{-3.87}^{+4.48}$ & $0.187_{-0.139}^{+0.355}$ \\[2pt]
NGC 5139 & \,$0.9\pm0.25\!\!\!$ & $12.3\pm0.3$ & (8) & 1.04 (D90) & 
$13.29_{-2.08}^{+2.55}$ & $0.200_{-0.127}^{+0.266}$ \\[2pt]
& 1.2 & $13.9^{+6.5}_{-4.5}$ & (6) & 0.823 (H10) $\!$ & 
$11.59_{-1.78}^{+2.20}$ & $0.199_{-0.115}^{+0.206}$ \\[2pt]
\hline\\*[-8pt]
NGC 6304 & 2.66 & $12.1^{+6.6}_{-4.8}$ & (9) & $3.46$ (G13) & 
$11.62_{-2.03}^{+2.80}$ & $0.212_{-0.113}^{+0.255}$ \\[2pt]
& & & & 2.66 (D90) & 
$9.54_{-1.66}^{+2.13}$ & $0.212_{-0.104}^{+0.141}$ \\[2pt]
& & & & 3.70 (H10) $\!$ & 
$12.54_{-2.33}^{+2.81}$ & $0.245_{-0.146}^{+0.258}$ \\
\enddata
\tablecomments{$N_H$ and $R_\infty$ values in columns
  2 and 3 are those quoted by the references cited in column 4: (1)
  -- \cite{Servillat12}; (2) -- \cite{Becker03}; (3) --
  \cite{Guillot11a}; (4) -- \cite{Gendre03b}; (5) --
  \cite{Catuneanu13}; (6) -- \cite{Webb07}; (7) --
  \cite{Rutledge02b}; (8) -- \cite{Gendre03a}; (9) --
  \cite{Guillot09}. 
The first row of columns 5, 6, and 7 for each
  source contains $N_H$, $R_\infty$, and $z$ values from G13 but
  recomputed to include distance uncertainties to excluding $M-R$
  regions forbidden by the combination of causality and the minimum
  maximum mass constraints. In the second row, $N_H$ is set to
  Chandra Toolkit {\tt http://cxc.harvard.edu/toolkit/colden.jsp}
  values (labelled D90). The third row has $N_H$ values using the
  reddening measurements $E(B-V)$ from \citet{Harris10} with the
  correlation between $E(B-V)$ and $N_H$ as discussed in
  \citet{Predehl95} and updated in \citet{Guver09} (labelled H10).
  For the second and third rows, $R_\infty$ values are scaled
  according to Table \ref{tab:nh2} ($z$ is unchanged by this procedure), and
  $R_\infty$ and $z$ are further corrected to account for causality
  and the minimum maximum mass constraints. All uncertainties
  reflect 90\% confidence.}
\end{deluxetable}

\begin{figure}
\begin{center}
\includegraphics[width=.49\textwidth]{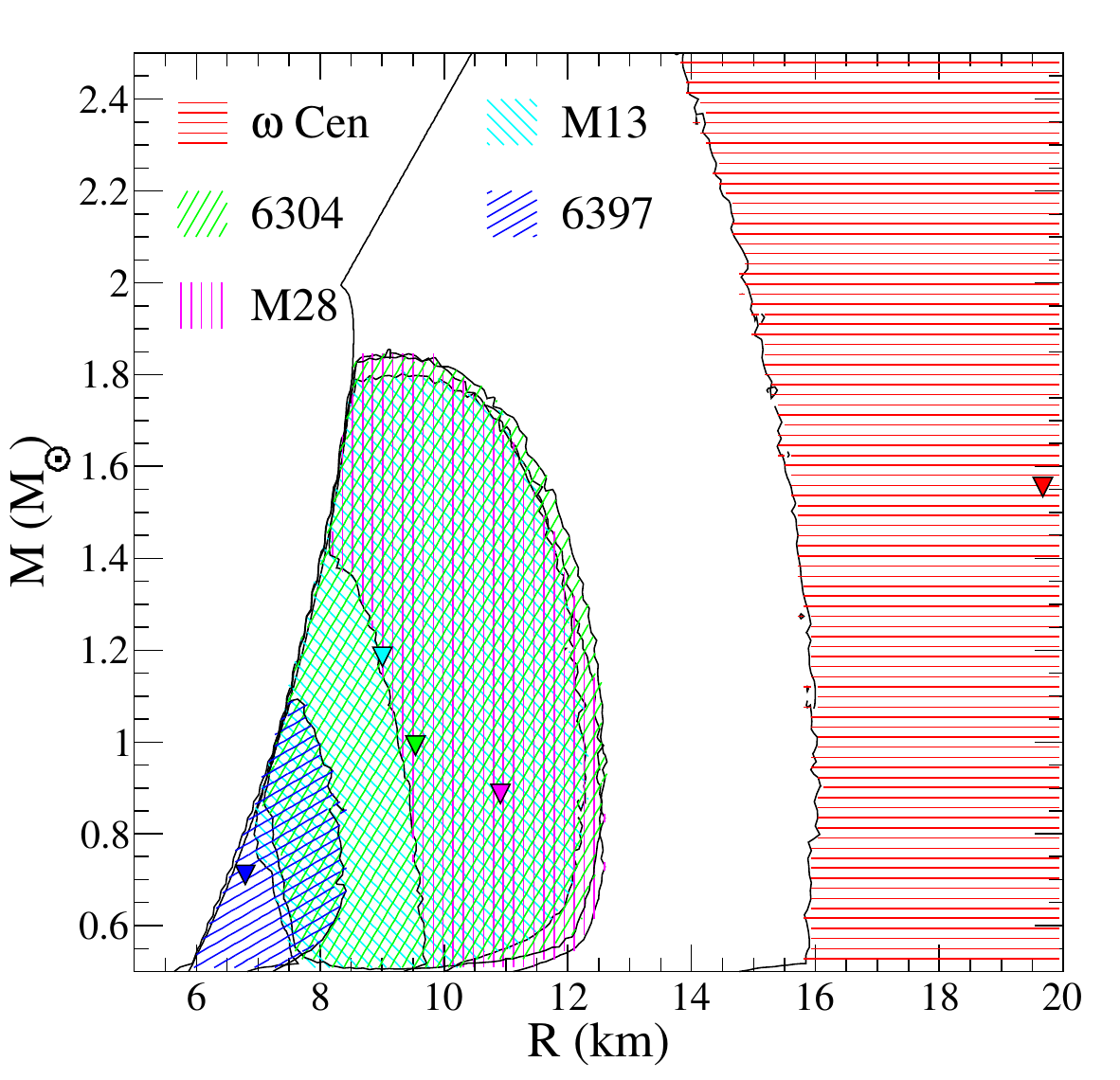}
\includegraphics[width=.49\textwidth]{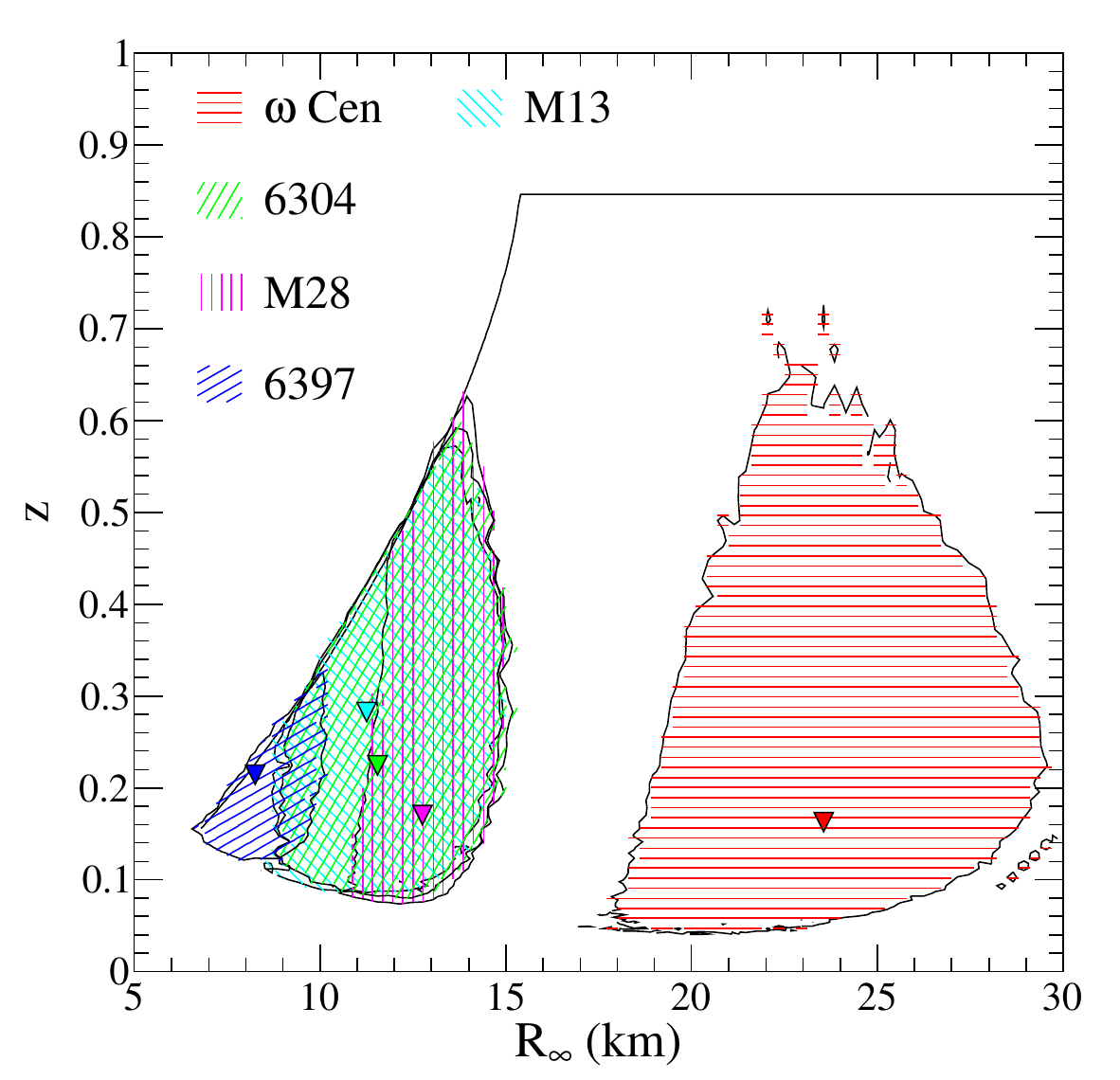}
\end{center}%
\vspace*{-.3in}
\caption{90\% uncertainty contours in $R_\infty$ and $z$ (left panel)
  and $M$ and $R$ (right panel) for QLMXBs studied by G13, including
  distance uncertainties. Triangles mark the most-favored values with
  ordering in $R_\infty$ or $R$ increasing for NGC 6397, M13, NGC
  6304, M28 and $\omega$ Cen. Contour fill patterns for each source
  are given in the legend. Regions to the left or above the solid
  lines are forbidden by the combination of causality and a
  2.0~M$_\odot$ lower limit to the neutron star maximum mass. }
 \label{fig:uncorr}
\end{figure}

We use the distances and uncertainties given in G13 and presented in
the second column of Table~\ref{tab:dist}, and also assume their
probability distributions can be modeled by Gaussians. Three of the
globular clusters, NGC 6397, M13 and $\omega$ Cen, have
dynamically-measured distances. M28 and NGC 6304 lack dynamical
distance estimates. However, other distance measurements exist. The
distance to NGC 6397 has been given as $2.52\pm0.1$ kpc in
\citet{Gratton03} and $2.54\pm0.07$ kpc in \citet{Hansen07}. We adopt
$2.53\pm0.1$ kpc as an alternate distance to the neutron star in NGC
6397. Also, \citet{Webb07} and \citet{Catuneanu13} use $7.7\pm0.6$ kpc
as the distance to M13 , which we also adopt as an alternative
distance. These alternatives are given in the third column of
Table~\ref{tab:dist}. Together with the values from the second column
for the other three sources, this set will be referred to as ``Alt''.

It is worth pointing out that the \citet{Gratton03} and
\citet{Hansen07} distances for NGC 6397, which agree with a
determination of 2.3 kpc from \citet{Harris10}, are 28\% larger than
what G13 used and differ by much more than their quoted errors. Also,
G13's distance to M13 is 18\% larger than those employed by
\citet{Webb07} and \citet{Catuneanu13}. These differences have an
impact on our results. Thus, there is a strong argument to also employ
a set of distance estimates obtained with similar systematics. Among
the largest analysis of globular cluster distances conducted in a
fairly homogeneous fashion is that of \citet{Harris10}. We arbitrarily
choose to combine these with the uncertainties from G13 and refer to
this set as H10 which appears in column four of Table~\ref{tab:dist}.
Because the inferred $M$ and $R$ coordinates of the distributions
scale linearly with $D$, $z$ is unaffected by the distance scale. In
the work below, we rescale and resample the $R_\infty\propto D$ and
$z$ coordinates of the distributions of G13 to obtain additional
probability distributions and then translate these to $(R,M)$
distributions.

Previous works have also made different assumptions about the extent
of X-ray absorption between the source and the observer. It has become
commonplace to characterize the magnitude of the X-ray absorption by a
single parameter, the ``equivalent hydrogen column density'' (we
denote the column density in units of $10^{21}$ atoms cm$^{-2}$ as
$N_{H21}$), which is the column density of atomic hydrogen which would
most closely replicate the net obscuration of
the spectrum including the effects of heavy elements in the
intervening material. G13 determined $N_H$ together with the
neutron star properties from spectral modeling, but the inferred
  $N_H$ values do not always agree with the value from HI maps summarized by \cite{Dickey90}.
For each source, the first row in column 5 in Table \ref{tab:nh} gives
the G13 determination, while the second row contains values of $N_H$
according to the Chandra Proposal Planning Toolkit {\tt
  http://cxc.harvard.edu/toolkit/colden.jsp}, hereafter referred to as
CPPT. The CPPT lists \cite{Dickey90} as its primary source, so we
refer to these alternative equivalent hydrogen column densities as
D90. We note that the values for $N_{H21}$ for M28, M13 and $\omega$
Cen from the CPPT slightly differ from those G13 attributed to
\cite{Dickey90} (G13 gives 2.4, 0.11 and 0.9, respectively, for these
sources). There are differences of up to 10-15\% among the $N_H$
values given from HEARSAC [{\tt
    http://heasarc.gsfc.nasa.gov/docs/tools.html}], CPPT, and the two
HI surveys, \citet{Dickey90} and \citet{Kalberla05}. We do not attempt
to completely resolve these differences but choose a range of
different $N_H$ values to systematically characterize the associated
uncertainty.

Independent HI surveys are not
necessarily good estimators of the true X-ray absorption. They can be
overestimates near the plane of the galaxy where they include
absorption from regions beyond the cluster. They do not include
molecular H$_{2}$, ionized H, and heavier elements, all of which 
can contribute to the absorption. Finally, the surveys do not always
have sufficient spatial resolution to resolve the true column in the
direction of the QLMXB and cannot account for absorption intrinsic to
the globular cluster. To circumvent these problems, an approach often
used is to take the HI surveys as a first guess, and improve that
guess where possible using either reddening measurements or direct
measurements of extinction from spectral fitting. As with the distance
measurements, it is therefore useful to compare with results using
uniform systematics, so we employ reddening measurements $E(B-V)$
\citep{Harris10} together with the correlation between $E(B-V)$ and
X-ray-measured $N_H$ as discussed in \citet{Predehl95} and updated in
\citet{Guver09}. These $N_H$ values are referred to as H10 in Table
\ref{tab:nh}. It is interesting that the H10 values for $N_H$ are
closer to the G13 values than are the D90 values for all the sources
except for $\omega$ Cen.

The probability distributions in $z-R_\infty$ space, which can be
  inferred from the $M-R$ results for the individual fits to each
  source in G13, tend to be banana-shaped, with the long axes
approximately characterized by a fixed value of $R_\infty$. It is
therefore useful to use $z-R_\infty$ distributions rather than $M-R$
distributions. The 90\% confidence intervals for the fits of G13 of
$R$, $M$, $R_\infty$ and $z$ are shown in Fig. \ref{fig:uncorr}. An
apparent feature of Fig. \ref{fig:uncorr} is that the most-probable
values of $R$ and $R_\infty$ span wide ranges so that $M$ and $R$ for
the 5 sources do not appear to follow traditional $M-R$ curves for
baryonic stars. In fact, it seems that $M$ and $R$ are roughly
linearly related. This implies these sources have similar redshifts,
although the uncertainties in $z$ for each source makes this difficult
to quantify.

It should also be noted that significant amounts of four of the five
probability regions, especially for NGC 6397, are forbidden by the
combination of causality (i.e., restricting the speed of sound to be
less than the speed of light), general relativity, and neutron star
mass observations. The minimum maximum neutron star mass ($\hat M$) is
$\hat M =2.0~M_\odot$, as implied by the mass $1.97 \pm 0.04$ of
pulsar PSR J1614-2230~\citep{Demorest10} and the mass $2.01 \pm 0.04$
of PSR J1614-2230~\citep{Antoniadis13}. The combination of general
relativity and causality alone imposes the restrictions
$R>2.83~GM/c^2$ and $z<0.847$ \citep{Lattimer07}, but the minimum
radius limit is larger for $M<\hat M$ \citep{Lattimer12}. This follows
from the conjecture of \citet{Koranda97} that the most compact
configurations are formed with the 'maximally compact' EOS
$p=\epsilon-\epsilon_0$ with the single parameter $\epsilon_0$. The
maximum neutron star mass for this EOS is given by $M_{\rm
  max}\simeq4.1\sqrt{\epsilon_s/\epsilon_0}~M_\odot$, where
$\epsilon_s\simeq150$ MeV is the energy density at the nuclear
saturation density ($\rho_s=2.7\cdot10^{14}$ g cm$^{-3}$). Using
$M_{max}>\hat M=2.0~M_\odot$, one finds $\epsilon_0\le4.2\epsilon_s$.
The curved portion of the combined maximum mass--causality constraint
is the $M-R$ curve of the maximally compact EOS with
$\epsilon_0=4.2\epsilon_s$~\citep{Lattimer05}. The most-probable
values and confidence ranges obtained by G13 for $R_\infty$ and $z$,
after excluding regions of $R_\infty-z$ space ruled out by causality
and $\hat M=2.0~M_\odot$, are given in the first row of columns 6 and 7
of Table~\ref{tab:nh} for each source. The values attributed to G13 in
Table~\ref{tab:nh} differ slightly from those of Table 4 in G13
because of these exclusions. This is most significant for the neutron
star NGC 6397 where about 60\% of the region is ruled out by causality
and similarly about 10\% of the region for the neutron star in M13.

G13 noted that the $N_H$ values they determined for their two most
extreme sources in terms of estimated radii (NGC 6397 and $\omega$
Cen) are significantly different from those of D90. G13 further noted
that varying the column density for heavily absorbed sources leads to
large changes in estimated radii. As an exercise, G13 recomputed
spectral fits for NGC 6397 and $\omega$ Cen fixing their $N_H$ values
to those of D90. The new values of $R_\infty$ were found to scale
approximately proportionately with the assumed values of $N_H$. We
note, as a result, the new values of $R_\infty$ G13 obtained for these
sources clustered near the values found for the other three sources.
This suggests that a more consistent picture might emerge if $N_H$ is
determined from independent HI surveys rather than from X-ray fitting.

This work aims to understand how different assumptions about the
atmospheric composition, distances, and X-ray absorption affect the
individual neutron star mass and radius determinations by developing
an analytical model of the neutron star atmosphere that can probe
these dependencies. This model is described and calibrated in \S
\ref{Sec:Col}. We also explore the sensitivity of inferred $M-R$
distributions to changes in surface gravity and redshift. In \S
\ref{Sec:altint}, employing various alternative assumptions, we
compute different $M-R$ distributions for QLMXBs studied by G13 and
explore the consequences for the neutron star mass-radius curve and
the EOS using the Bayesian techniques developed in
\citep{Steiner10,Steiner13}. We implicitly incorporate the constraints
of causality, the observed minimum maximum mass, well-established
properties of the hadronic neutron star crust, and the relationship
between the EOS and $M$ and $R$ provided by the
Tolman-Oppenheimer-Volkov (TOV) structure equations.

\section{DEPENDENCE ON $N_H$ AND COMPOSITION}        
\label{Sec:Col}

As already noted above, the determinations of $M$
and $R$ are very sensitive to assumptions for $D$ and $N_H$. The
direction of the change in inferred values of $R_\infty$ is as
expected: increasing $N_H$ has the same qualitative effect as
increasing $D$, for which inferred values of $M$ and $R$, as well as
$R_\infty$, scale with the assumed distance. G13 found that the effect
is quite large and that $R_\infty$ increases approximately
proportionally with the assumed value of $N_H$. In the case of NGC
6397, G13 found that an increase in $N_H$ by a factor of 1.55 produces
an increase in $R_\infty$ of a factor 1.42, while in the case of
$\omega$ Cen, a decrease in $N_H$ by a factor of 0.49 produced a
decrease in $R_\infty$ by a factor 0.51 .

The analytical model we present here is based on the absorptive
effects of atomic hydrogen on the spectrum as a simple model for X-ray
absorption. In general, the true X-ray absorption is dominated by
heavier elements and molecular hydrogen. We use the hydrogen column
density as a proxy for the net absorption. We correct for this
distinction at the end by modifying our analytical model so that it
quantitatively reproduces the behavior observed by G13 in response to
$N_H$ changes. This procedure is at best qualitatively
correct, but it enables us to transparently examine how absorption can
affect the inferred mass and radius distributions.

\subsection{Blackbody Atmosphere}
\label{Sec:bb}
Because X-ray absorption by interstellar H is frequency-dependent, the
inferred effective temperature $T_{\rm eff}$ depends on
assumptions concerning $N_H$. It is straightforward to estimate the
magnitudes of changes in the inferred $T_{\mathrm{eff}}$ and
$R_\infty$ values for different assumptions concerning $N_H$. As a
first illustration, we employ a blackbody model. A simplification the
blackbody model provides is that the emergent spectrum is independent
of gravity or redshift for a given total flux and observed peak
energy. We will subsequently explore the effects of gravity when
considering hydrogen and helium atmospheres.

The observed energy dependence of the flux from an absorbed blackbody
with an effective temperature $T_{\mathrm{eff}}$ obeys
\be\label{planck}
F(E,T_{\mathrm{eff}},N_H)=\alpha E^3\left({e^{-bN_{H21}/E^{8/3}}\over e^{E/kT_{\mathrm{eff}}}-1}\right),
\ee
where $\alpha$ is a constant and $b\simeq0.16$ keV$^{8/3}$. The term
involving $N_H$ represents the approximate effects of
absorption~\citep{Wilms00}. For a given $T_{\mathrm{eff}}$, the
maximum flux occurs at $E_{0}$ where $dF/dE=0$, or
\begin{equation}\label{peak}
3-(E_{0}/kT_{\mathrm{eff}})(1-e^{-E_{0}/kT_{\mathrm{eff}}})^{-1}
+(8/3)bN_{H21}E_{0}^{-8/3}=0.
\end{equation}
The solution of this equation leads to $E_0\simge3kT_{\mathrm{eff}}$
in general, so that the small exponential term in parentheses can be
neglected, leading to
\be\label{peak1}
E_0\simeq kT_{\mathrm{eff}}\left({9+8bN_{H21}E_0^{-8/3}\over3}\right).
\ee
To compare the effect of changing the amount of absorption on the
inferred radius, we assume the total observed flux and the peak energy
$E_{0}$ are held fixed as $N_H$ is changed. Keeping $E_0$ fixed,
changing the hydrogen column density from $N_1\equiv N_{1,H21}$ to
$N_2\equiv N_{2,H21}$ will alter the inferred effective temperature
from $T_{\mathrm{eff},1}$ to $T_{\mathrm{eff},2}$:
\be\label{eff}
{T_{\mathrm{eff},2}\over
  T_{\mathrm{eff},1}}\simeq{9E_{0}^{8/3}+8bN_{1}\over9E_{0}^{8/3}+8bN_{2}},
\ee 
where we again neglected the factor $1-e^{-E_{0}/kT_{\mathrm{eff}}}$ from Eq.
(\ref{peak}). Therefore, the effective temperature will decrease with
an increase in assumed column density.

The total observed flux is $(R/D)^2\int_{E_L}^{E_U}F(E)dE$, where
$E_L\sim0.3$ keV and $E_H\simeq10$ keV represent the low- and
high-energy cutoffs of the X-ray detector response function.
Therefore, conservation of the observed flux leads to the following
relation between the inferred neutron star radii in the two cases:
\be\label{rrat}
\left({R_2\over R_1}\right)^2={\int_{E_L}^{E_U}F(E,T_{\mathrm{eff},1},N_{1})
dE\over\int_{E_L}^{E_U}F(E,T_{\mathrm{eff},2},N_{2})dE}.
\ee%
The integrals in Eq. (\ref{rrat}) can be approximately evaluated by
the method of steepest descent, in which the lower and upper
integration limits are extended to $-\infty$ and $\infty$,
respectively, and the integrand is replaced by a Gaussian centered at
$E_0$, which leads to
\be\label{rrat1}
\left({R_2\over R_1}\right)^2&\simeq&{F_1\over
F_2}\sqrt{F^{\prime\prime}_2 F_1\over F_2F^{\prime\prime}_1},
\ee
where $F_1\equiv F(E_{0},T_{\mathrm{eff},1},N_1)$ and $\prime\prime$ indicates a
second derivative with respect to energy evaluated at the
peak. Approximately, we have
\begin{eqnarray}\label{rrat2}
{F_1\over F_2}&\simeq&
\exp\left[{11b(N_{2}-N_{1})\over3E_{0}^{8/3}}\right],\cr
{F^{\prime\prime}_2F_1\over F_2F^{\prime\prime}_1}&\simeq&
{27E_{0}^{8/3}+88bN_{2}\over27E_{0}^{8/3}+88bN_{1}}.
\end{eqnarray}
An increase in $N_H$ necessarily leads to an increase in $R$, since
the values of both Equations (\ref{rrat2}) are greater than unity. As
a numerical example, for $T_{\mathrm{eff},1}=0.1$ keV, $N_{1}=0.9$ and $N_{2}=1.8$,
one finds $E_{0}\simeq0.52$ keV, $T_{\mathrm{eff},2}\simeq0.070$ keV, and
$R_2/R_1\simeq5.35$. The analytic expressions Eqs. (\ref{rrat1}) and
(\ref{rrat2}) are accurate, compared to the full expression Equation
(\ref{rrat}), to better than 1\%. This radius ratio is to be compared
to $R_{2\infty}/R_{1\infty}\simeq2$ found by G13 for hydrogen
atmosphere models in the case of $\omega$ Cen (Table \ref{tab:nh}).

\subsection{Hydrogen Atmosphere}        
\label{Sec:hydrogen}

We next consider the case of a hydrogen atmosphere. The observed
spectrum can be approximated by
\begin{equation}\label{planck1}
F(E,T_{\mathrm{eff}},N_H)=\alpha^\prime(T_{\mathrm{eff}}) E^3(e^{\beta(E/T_{\mathrm{eff}})^p}-1)^{-1}e^{-bN_{H21}/E^{8/3}},
\end{equation}
where $\alpha^\prime$ now depends on $T_{\mathrm{eff}}$, $\beta\simeq 1.234$, and
$p\simeq5/7$. The value for $p$ comes from considerations of the
competition between electron scattering and free-free absorption in a
gray atmosphere, and $\beta$ is obtained by fitting realistic hydrogen
atmosphere models \citep{Romani87,Zavlin96} with temperatures near 0.1
keV. We are grateful to Ed Brown for bringing this approximation to
our attention. The effect of $p<1$ is to broaden the energy
distribution, so that there is a larger fraction of high-energy
photons emitted compared to a blackbody. Justification for the value
of $p$ can be found in Appendix A.

Assuming the observed peak energy is fixed, the effective temperatures
for two different assumed column densities are now related by
\be\label{eff1}
{T_{\mathrm{eff},2}\over
  T_{\mathrm{eff},1}}\simeq\left({9E_{0}^{8/3}+8bN_{1}
\over9E_{0}^{8/3}+8bN_{2}}\right)^{1/p},
\ee 
again ignoring small exponential factors. One can find the ratio of
inferred radii using Eq. (\ref{rrat}), where the approximations Eq.
(\ref{rrat1}) and
\begin{eqnarray}\label{rrat3}
{F_1\over F_2}&\simeq&
\left({T_{\mathrm{eff},1}\over T_{\mathrm{eff,2}}}\right)^{0.2}
\exp\left[\left({8\over3p}+1\right)
b{N_{2}-N_{1}\over E_{0}^{8/3}}\right],\cr
{F^{\prime\prime}_2F_1\over F_2F^{\prime\prime}_1}&\simeq&
{27pE_0^{8/3}+8bN_2(8+3p)\over27pE_0^{8/3}+8bN_1(8+3p)},\hspace*{1cm}
\end{eqnarray}
are accurate to about 1\% compared to the full expressions. For values
of $T_{\mathrm{eff},1}=0.1$ keV, $N_1=0.9$ and $N_2=1.8$, we now find
that $R_2/R_1\simeq2.24$, considerably smaller than the blackbody
result, and near the result reported by G13 for realistic atmosphere
simulations. The primary reason for a smaller radius increase,
relative to the blackbody case, is that $p<1$. It is found in this
case that $E_0=0.78$ keV and $T_{\mathrm{eff,2}}=0.078$ keV.

\begin{figure}
\begin{center}
\hspace*{-.5cm}\includegraphics[width=.55\textwidth]{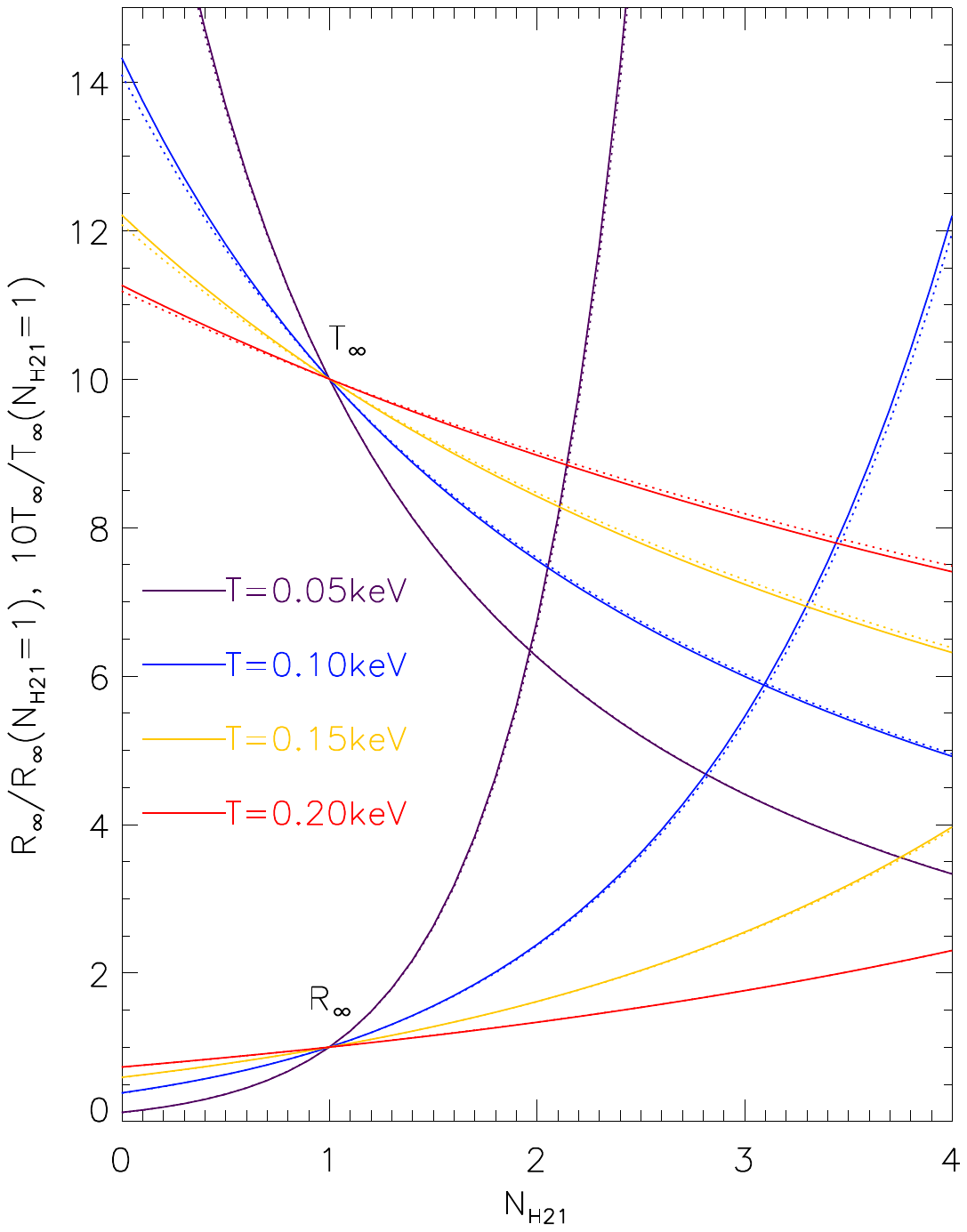}
\end{center}%
\caption{Radii (lower curves) and effective
temperatures (upper curves) for hydrogen atmospheres as functions of
column densities. Results are shown as ratios relative to the values
obtained for the indicated base effective temperature and column
density ($N_{H21}=1.0$). Solid lines show numerical results and
dotted lines show the results of the analytical expressions
discussed in the text.
\label{radrat}}
\end{figure}

The increase of inferred $R$ with increasing $N_H$ is quite sensitive
to the temperature. Fig. \ref{radrat} shows these changes for
different base effective temperatures. The dotted curves show the
approximate analytic results using Eqs. (\ref{rrat1}), (\ref{eff1}),
and (\ref{rrat3}). It can be seen that, despite the nearly linear
relation between $N_H$ and $R_\infty$ deduced by G13 for NGC 6397 and
$\omega$ Cen, the linearity is accidental. In summary, the effects of
appreciably changing $N_H$ are magnified by relatively low effective
temperatures, and furthermore are large only for highly obscured
sources (i.e., for $N_{H21}\simge1$).

\subsection{Helium Atmosphere}
\label{Sec:helium}
The case of a helium atmosphere follows similarly to that of the
hydrogen case. As in the case of hydrogen, we expect that electron
scattering opacities dominates the mean free path so the effective
exponent $p=5/7$ can be assumed for the spectral shape.
\cite{Suleimanov11} computed an array of spectral energy distributions
of hydrogen and helium atmospheres as functions of temperature and
gravity. Although these tables do not extend to temperatures below
about 0.3 keV, comparison shows, for a given gravity, that helium
atmospheres with temperatures approximately 13\% smaller than hydrogen
atmospheres have nearly identical energy distributions and peak
energies. Therefore, a simple approximation to a helium atmosphere
would be that of Eq.~(\ref{planck1}) with $\beta\simeq1.24$ instead of
1.35. In this case, fitting an observed spectrum with a helium, rather
than a hydrogen, atmosphere, and requiring that the peak in the energy
distribution remain unaltered, leads to an inferred temperature
decrease by 13\% and a corresponding inferred radius increase by about
28\%.

For comparison, \cite{Catuneanu13} analyzed data from Chandra, XMM and
ROSAT for the M13 QLMXB using hydrogen and helium atmospheres. They
determined that fits with helium atmospheres resulted in increases in
radiation radii by a factor of approximately 1.25. However, because
the helium fit involved a reduction in surface gravity, and as we show
below, the effective value of $\beta$ depends slightly on the assumed
gravity, part of this radius change can be attributed to gravity
effects. In contrast, \cite{Servillat12} analyzed data for the M28
QLMXB and reported an increase in $R_\infty$ of approximately 50\% for
a helium instead of hydrogen atmosphere. Their helium fit also
resulted in a lower gravity than the hydrogen fit. In addition, it
could be expected that the relative inferred radius change could also
be a function of $N_H$. Exploring this in detail is beyond the scope
of our analytic model. We simply assume, for the calculations
described below, that a change in atmospheric composition, at fixed
gravity, from hydrogen to helium results in a radiation radius
increase of approximately 33\%. Therefore, we expect our simulations
describing the effects of considering He atmospheres to be even more
qualitative than those of varying distance or $N_H$. At fixed
redshift, any inferred radius change will necessarily result in a
gravity change.

\subsection{Dependence on Gravity and Redshift}        
\label{Sec:redshift}

To lowest order, hydrogen and helium atmospheres have a relatively
weak dependence on gravity. The effects of gravity can be modeled as a
small modification to the parameter $\beta$ in Eq. (\ref{planck1}).
Without any temperature or gravity dependence in either $\beta$ or
$\alpha^\prime$, the total integrated flux would be proportional to
$R^2T^4$, and after applying a source redshift, the observed flux
becomes proportional to $R_\infty^2T_\infty^4$ which is the
redshift-independent blackbody result. The net dependence on gravity
stemming from gravity and temperature dependencies in $\alpha^\prime$
and $\beta$ allow, in principle, gravity or redshift information to be
deduced from the spectrum.

The array of atmospheres computed by \cite{Suleimanov11} allows an
estimate of the effect of gravity to be made. For the lowest
temperatures they model, the effect of gravity on the spectral
distributions for hydrogen atmospheres can be approximated with a
gravity-dependent $\beta$ parameter:
\begin{equation}\label{betag}
\beta\simeq1.35\left({g\over10^{14.3}{\rm~cm}^2{\rm~s}^{-1}}\right)^{0.1}.
\end{equation}
A similar change in the $\beta$ factor for helium atmospheres is
found. This gravity dependence is sufficiently weak that, to lowest
order, it is safely ignored for the semi-analytic approximations
discussed below. In support of this thesis, we note that in the
comparison made by \cite{Servillat12}, replacing hydrogen by helium
resulted in an increase in $R_\infty$ of about 50\% but a reduction in
$1+z$ of only 4\%.

In any case, other information in addition to the peak energy and
total flux, such as higher spectral moments, need to be considered in
order to be able to obtain redshift constraints from spectral fitting.
This accounts for the relatively large uncertainties in redshift
estimated by G13 (Table \ref{tab:nh}). We will use the original values for
  $z$ as computed by G13 when varying $N_H$. Nevertheless, these
approximations make our analysis qualitative in nature, and moderate
changes to inferred neutron star properties can be expected from a
more sophisticated analysis.

\subsection{Calibration of the Semi-analytic Model}
\label{Sec:final}

We now proceed to test and calibrate the predictions of the analytic
model for variations in absorption by comparing to the atmosphere
simulations performed by G13. G13 performed this exercise for the
neutron stars in NGC 6397 and $\omega$ Cen. They compared the optimum
values of $R_\infty$ obtained for $N_H$ values determined through
spectral fitting and the $N_H$ values they attributed to
\citet{Dickey90}. We summarize this exercise in rows 7 and 8 of
columns 2-6 of Table \ref{tab:comp}, where the subscript 1 refers to
G13's $N_H$ value and 2 refers to values they attributed to
\citet{Dickey90} for the sources in NGC 6397 and $\omega$ Cen. Columns
7 and 8 of these rows show the results of the analytic model.

\begin{deluxetable}{l|ccccc|ccc}
\tablecaption{Dependence of Inferred Radii on $N_H$ for H Atmospheres
}
\tablewidth{0pt}
\tablehead{
\colhead{Source} & \colhead{$T_{\mathrm{eff},1}$ (eV)} & \colhead{$N_1$} & 
\colhead{$N_2$} & \colhead{$T_{\mathrm{eff,2}}$ (eV)} & 
 \colhead{$R_2/R_1$} & \colhead{$T_{\mathrm{eff,2}}$ (eV)} &
\colhead{$R_2/R_1$} & \colhead{$(R_2/ R_1)^{2/3}$}\\
& & & & & &\multicolumn{3}{c}{Semi-analytic model}
}
\startdata
example (see text) & 100 & 0.9 & 1.8 & --- &  --- & 78 & 2.3 & 1.8 \\
4U 2129+49 [1]& $80\pm21$ & 2.8 & 1.7 & $100^{+25}_{-20}$ & 
$0.42\pm0.13$ & $100\pm21$ & $0.47^{+0.08}_{-0.10}$ & $0.60^{+0.07}_{-0.08}$ \\[2pt]
4U 1608-522 [2]& $170\pm30$ & 8.0 & 15.0 & $105^{+20}_{-16}$ & 
$5.3\pm3.6$ & $125\pm27$ & $2.9^{+0.9}_{-0.5}$ & $2.0^{+0.4}_{-0.2}$ \\[2pt]
Cen X-4 [2]& $100\pm12$ & 0.55 & 2.0 & $63^{+7}_{-8}$ & 
$4.1\pm1.5$ & $60\pm11$ & $5.3^{+2.1}_{-1.2}$ & $3.0^{+0.8}_{-0.4}$ \\[2pt]
Aql X-1 [2]& $250\pm25$ & 2.0 & 4.0 & $200\pm25$ & 
$1.7\pm0.6$ & $223\pm27$ & $1.4\pm0.1$ & $1.24^{+0.06}_{-0.04}$ \\[2pt]
Aql X-1 [2]& $200\pm25$ & 4.0 & 8.0 & $140\pm23$ & 
$2.9^{+1.2}_{-1.8}$ & $159\pm25$ & $2.0^{+0.3}_{-0.2}$ & $1.6\pm0.1$ \\[2pt]
NGC 6397 [3]& $76^{+15}_{-6}$ & 0.96 & 1.4 & 64 & 
$1.4\pm0.1$ & $64^{+15}_{-6}$ & $1.8^{+0.1}_{-0.2}$ & $1.5\pm0.1$ \\[2pt]
$\omega$ Cen[3] & $64^{13}_{-5}$ & 1.82 & 0.9 & 87 &
$0.51\pm0.10$ & $88^{+14}_{-5}$ &  $0.36^{+0.07}_{-0.03}$ & 
$0.51^{+0.06}_{-0.03}$ \\[2pt]
\enddata
\tablecomments{The example is described in the text. Columns 2--6 are
  taken from the indicated publications: [1] \cite{Rutledge00}, [2]
  \cite{Rutledge99}, [3] G13. Columns 7--9 are semi-analytic model
  results.\label{tab:comp}}
\end{deluxetable}

Further indication that our approximate study qualitatively predicts
the effects of varying $N_H$ can be found in studies of
quiescent X-ray binaries in the field. For example, \cite{Rutledge00}
studied the effects of varying $N_H$ in spectral modeling of the
source 4U 2129+47 and \cite{Rutledge99} similarly studied the sources
4U 1608-522, Cen X-4 and Aql X-1, as summarized in Table
\ref{tab:comp}. It is interesting that the semi-analytic model
accurately predicts the changes in effective temperatures accompanying
variations in absorption. Predicted radius changes are much
less accurate, but given that 90\% confidence intervals for the
reported radius ratios can be greater than $\pm50$\%, the agreement is
satisfactory.

\begin{deluxetable}{lcccccccc}
\tablecaption{Scaling Factors from the Semi-analytic Model\label{tab:nh2}}
\tablehead{
\colhead{Source} &
\colhead{$N_{1}$} &
\colhead{$T_{\mathrm{eff},1}$ (eV)} &
\colhead{$R_{1\infty}$ (km)} &
\colhead{$N_H$ set} &
\colhead{$N_{2}$} &
\colhead{$T_{\mathrm{eff,2}}$ (eV)} &
\colhead{$R_{2}/R_{1}$} &
\colhead{$(R_{2}/R_{1})^{2/3}$} 
}
\startdata
M28 & 2.52 & 119 & $12.91^{+0.54}_{-0.61}$ & 
D90 & $1.89$ & 131 & 0.756 & 0.830 \\
&  &  &  & 
H10 & $2.74$ & 116 & 1.099 & 1.065 \\
\hline\\*[-.8pc]
NGC 6397 & 0.96 & 76 & $8.40^{+0.32}_{-0.32}$ & 
D90 & $1.4$ & 64 & 1.73 & 1.44  \\
&  &  &  & 
H10 & $1.23$ & 71 & 1.398 & 1.25 \\
\hline\\*[-.8pc]
M13 & 0.08 & 86 & $11.49^{+1.03}_{-0.97}$ & 
D90 & $0.145$ & 80 & 1.20 & 1.13 \\
&  &  &  & 
H10 & $0.137$ & 83 & 1.17 & 1.11 \\
\hline\\*[-.8pc]
$\omega$ Cen & 1.82 & 64 & $23.20^{+2.15}_{-2.08}$ & 
D90 & $1.04$ & 82 & 0.433 & 0.572 \\
&  &  &  & 
H10 & $0.823$ & 81 & 0.350 & 0.496 \\
\hline\\*[-.8pc]
NGC 6304 & 3.46 & 106 & $11.62^{+1.47}_{-1.26}$ & 
D90 & $2.66$ & 118 & 0.713 & 0.80 \\
&  &  &  & 
H10 & $3.70$ & 104 & 1.10 & 1.07 \\
\enddata
\tablecomments{Columns 2--4 are results from \cite{Guillot13}.  
Columns 6--9 show semi-analytic model results for the alternative $N_H$ sets:
D90, {\tt  http://cxc.harvard.edu/toolkit/colden.jsp}, in the first row; 
H10 \citep{Harris10} in the second row of each entry.
}
\end{deluxetable}

Our relatively simple representation of H atmospheres is seen to
overpredict the changes in radii due to variations in column density
when compared to the results of G13 for the sources in NGC 6397 and
$\omega$ Cen. This is not surprising, given that we have approximated
absorption as being due to H rather than to heavier elements. However,
it is clear that taking the 2/3 power of our predicted scaling
factors $R_{\infty2}/R_{\infty1}$ rather closely represents the
results of G13 for these sources, as displayed in the last column of
Table \ref{tab:comp}. In fact, this procedure still fits the other
radius ratios listed in Table \ref{tab:comp} at the 90\% confidence
level. Our final procedure for the generation of alternate $M-R$
probability distributions is therefore to rescale the $M$ and $R$
coordinates of the probability distributions shown in Figure
\ref{fig:uncorr} by the $2/3$ power of the amount predicted from
Equation (\ref{rrat}) for the $N_H$ values for each source from either
D90 or H10 (also using G13's values for $T_{\mathrm{eff},1}$). Our
analytical model does not allow us to predict the value $2/3$.
Nevertheless, our general conclusions are not sensitive to this power,
at least in the range of 1/2 to 3/2, since the ratios $N_2/N_1$ for
different models have no dominant trend.

\section{VARIATIONS OF X-RAY ABSORPTION, DISTANCE AND ATMOSPHERIC COMPOSITION}
\label{Sec:altint}

We use the final procedure described in \S \ref{Sec:final} to modify
the predicted radiation radii and $M-R$ probability distributions from
G13 for the D90 $N_H$ values (left panel of Fig.~\ref{fig:corr}), and
for the H10 $N_H$ values (right panel of Fig.~\ref{fig:corr}). (See
Appendix B for some additional details on how the correction is
applied for $\omega$ Cen.) The scaling factors are explicitly
displayed in Table \ref{tab:nh2} for the two alternative $N_H$
assumptions.

There are few discernible trends in the values of $N_H$ for the three
cases, except for the fact that the average $N_H$ value from D90 is
about 20\% smaller than that of G13 and about 20\% larger than that of
H10. The H10 values are larger than the G13 values in four of the five
sources; the D90 values are larger than the G13 values in two of the
five sources. As noted before, the H10 values are closer to the G13
values in all cases except for $\omega$ Cen. The average radius of the
five sources for the three $N_H$ sets are predicted to be within about
1 km of one another. The major differences between using the D90 and
H10 $N_H$ values occurs for M28 and NGC 6304 which have larger radii
in the H10 case. Indeed, for the D90 case, NGC 6304 appears to have 
relatively small values of  $M$ and $R$. Nevertheless,
Fig.~\ref{fig:corr} shows that, in either case D90 or H10, the new
radii implied by the adjusted $N_H$ values are more consistent than
the radii determined by G13 with the expectation that all neutron
stars have similar radii, as suggested by the results of
\citet{Steiner10} and \citet{Steiner13}.

\begin{figure}
\begin{center}
\includegraphics[width=.49\textwidth]{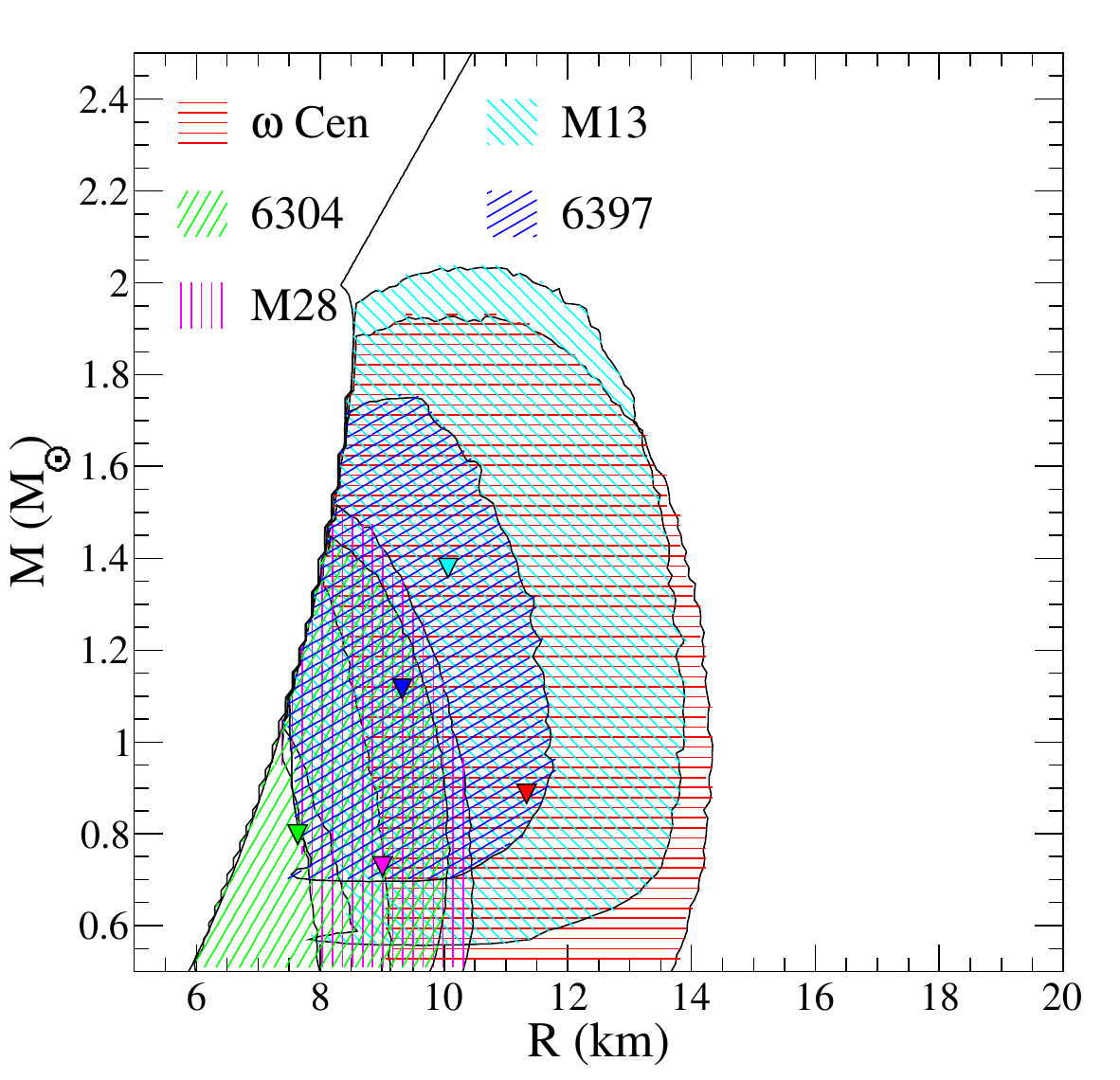}
\includegraphics[width=.49\textwidth]{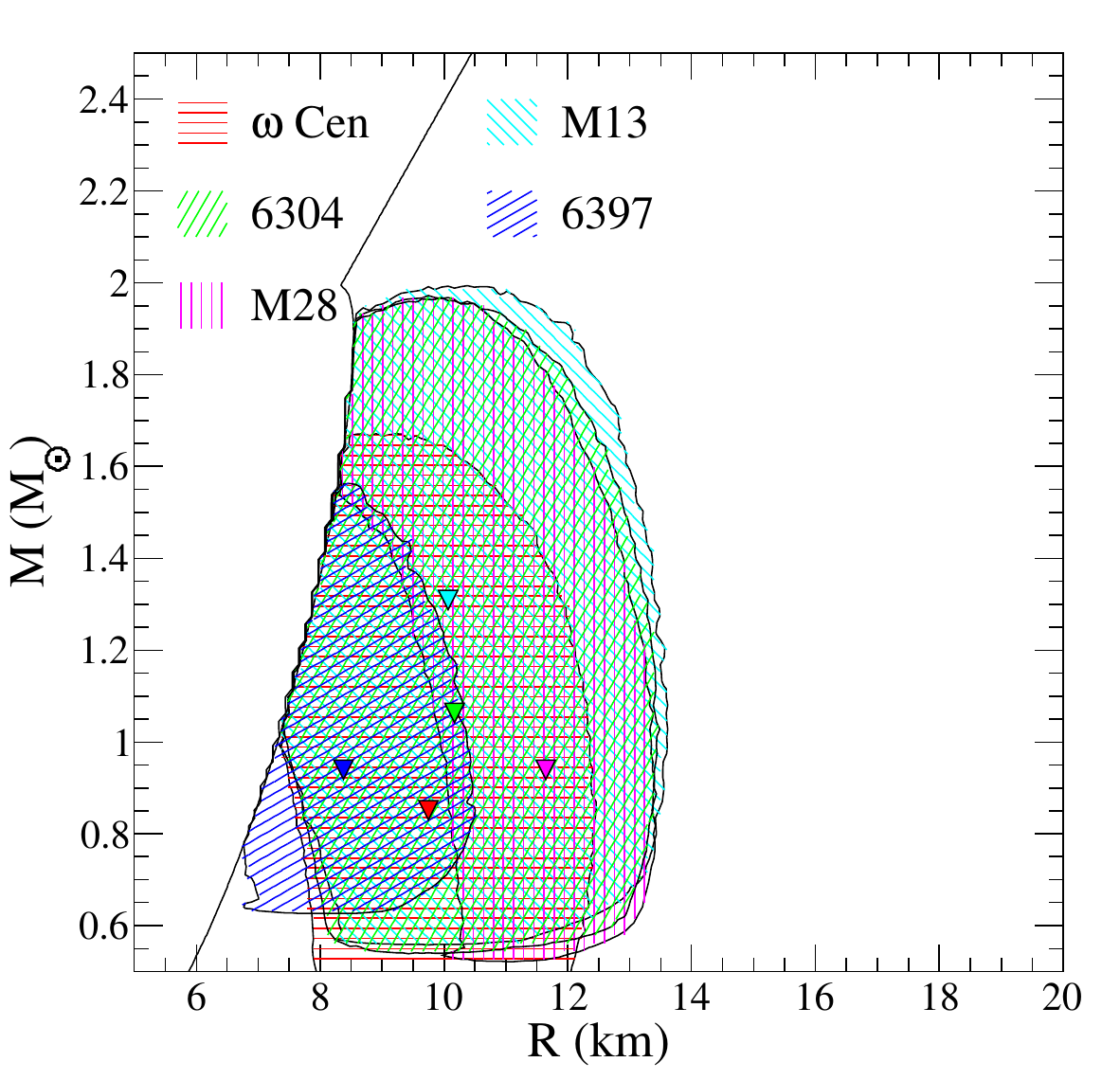}
\end{center}%
\caption{Similar to Fig. \ref{fig:uncorr}, but employing $N_H$ values
  from D90 (left panel) and H10 (right panel) and scaling the $M-R$
  probability distributions as described in the text.
 \label{fig:corr}}
\end{figure}

\begin{figure}
\begin{center}
\includegraphics[width=.49\textwidth]{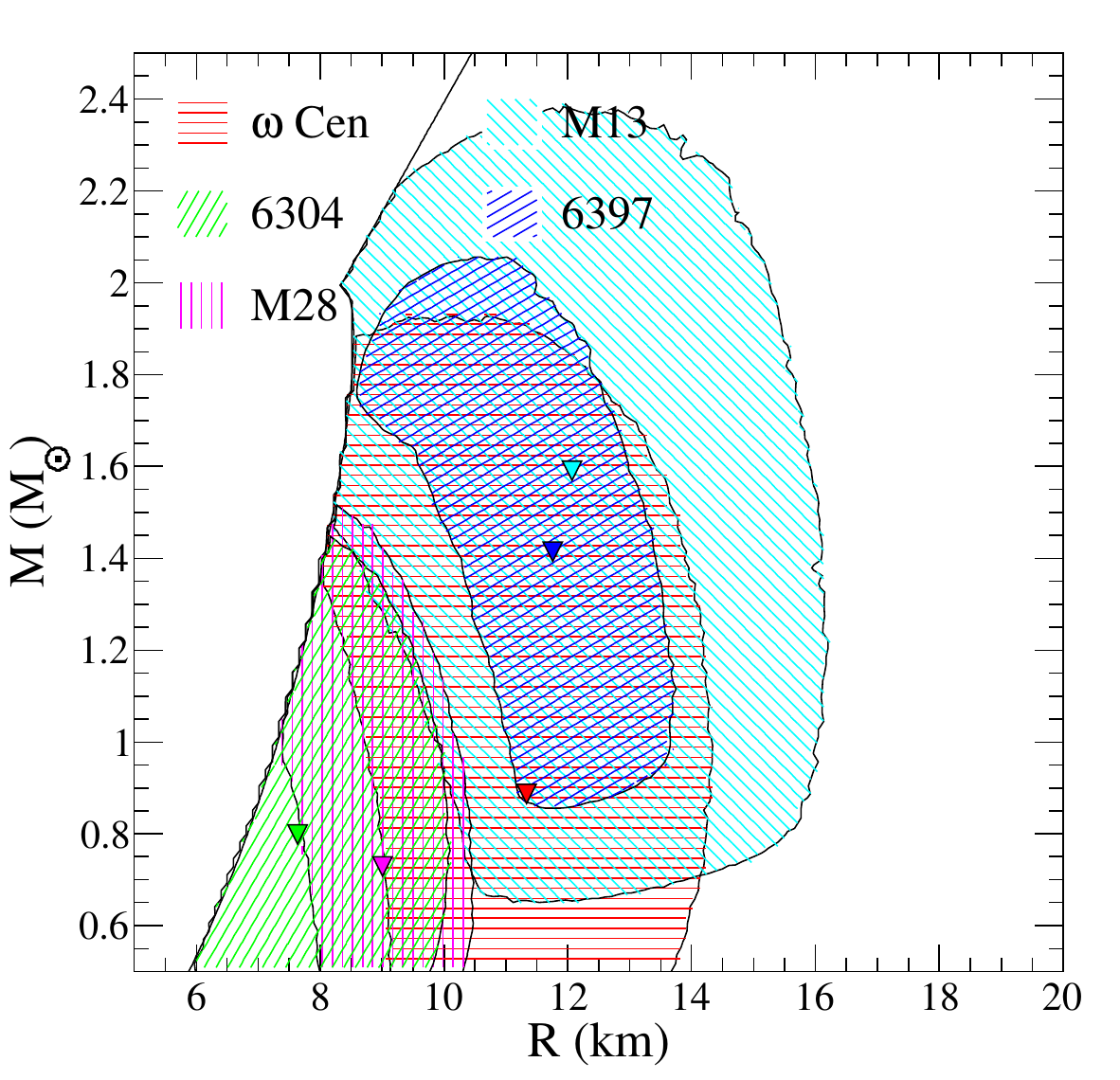}
\includegraphics[width=.49\textwidth]{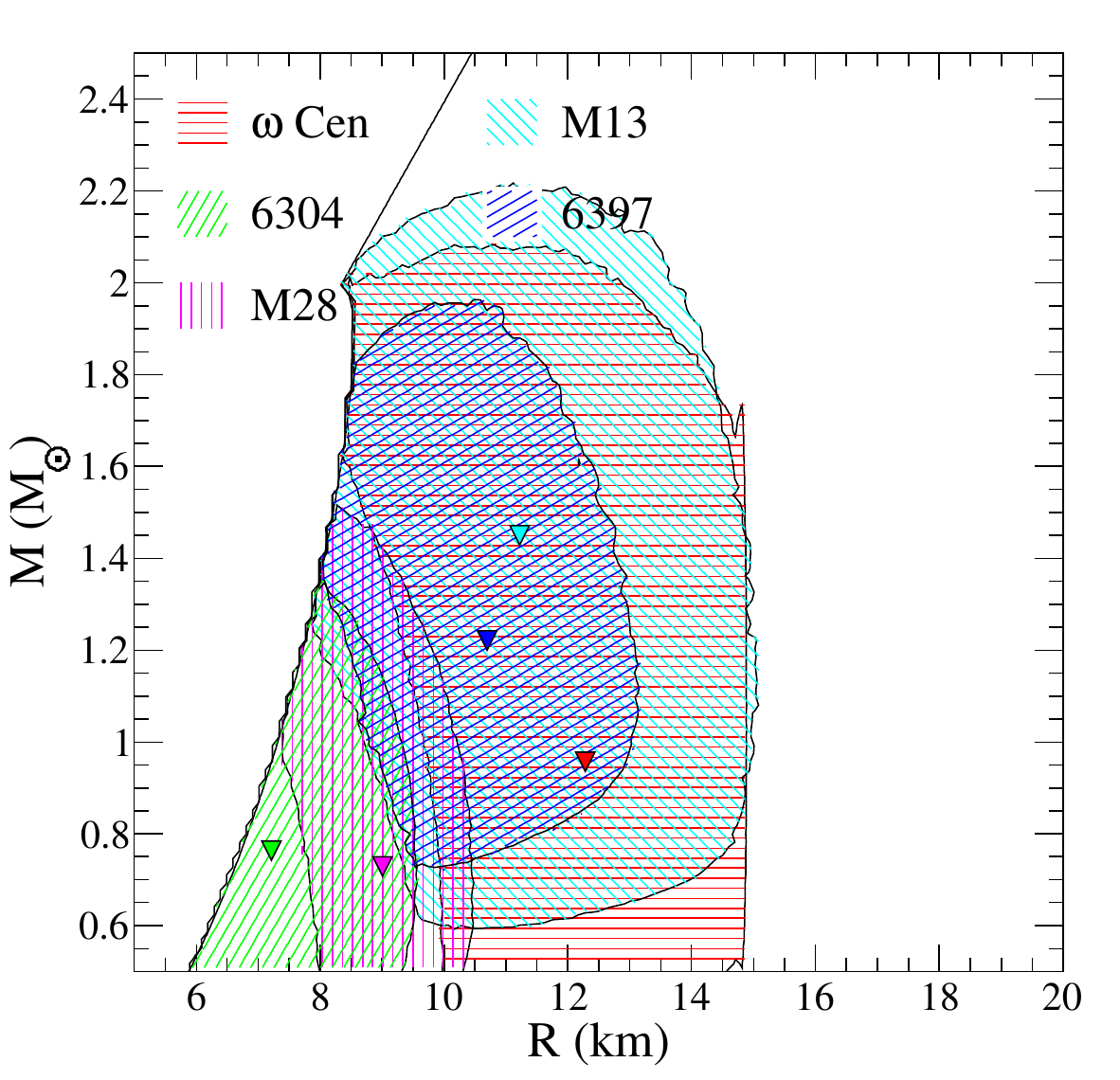}
\end{center}%
\caption{Similar to the left panel of Fig. \ref{fig:corr}, but
  employing the alternate (Alt) distances from the third column of Table
  \ref{tab:dist} (left panel) or the distances from H10 (right panel)
  and scaling the $M-R$ probability distributions as described in the
  text.
 \label{fig:distvar}}
\end{figure}

We also employ different distance measurements, and this variation is
studied in Fig.~\ref{fig:distvar}. These results assume the $N_H$
values from D90 and are thus to be compared with the left panel from
Fig.~\ref{fig:corr}. There is some variation in the implied values of
$R_{\infty}$, particularly for M13 and NGC 6397. Nevertheless, the
different distances give rise to smaller variations than do the different
values of $N_H$, and the $M-R$ probability regions generally overlap as
long as one of the alternative $N_H$ sets are used.

The radius uncertainties in the above analysis could be underestimated
because it is possible that one or more of these QLMXB sources have
helium, rather than hydrogen, atmospheres. Ultracompact binaries,
which have neutron stars that have accreted helium or carbon
atmospheres from white dwarf companions, are more likely to be found
in globular clusters than in the field \citep{Deloye04,Ivanova08}.
 Only in the cases of the neutron star in $\omega$ Cen
\citep{Haggard04}, X5 in 47 Tuc \citep{Heinke05} and several field
sources has it been confirmed that the atmosphere is hydrogen. Sources
with abnormally small inferred values of $R_\infty$ could have helium,
rather than hydrogen, atmospheres.

\begin{figure}
\begin{center}
\includegraphics[width=.49\textwidth]{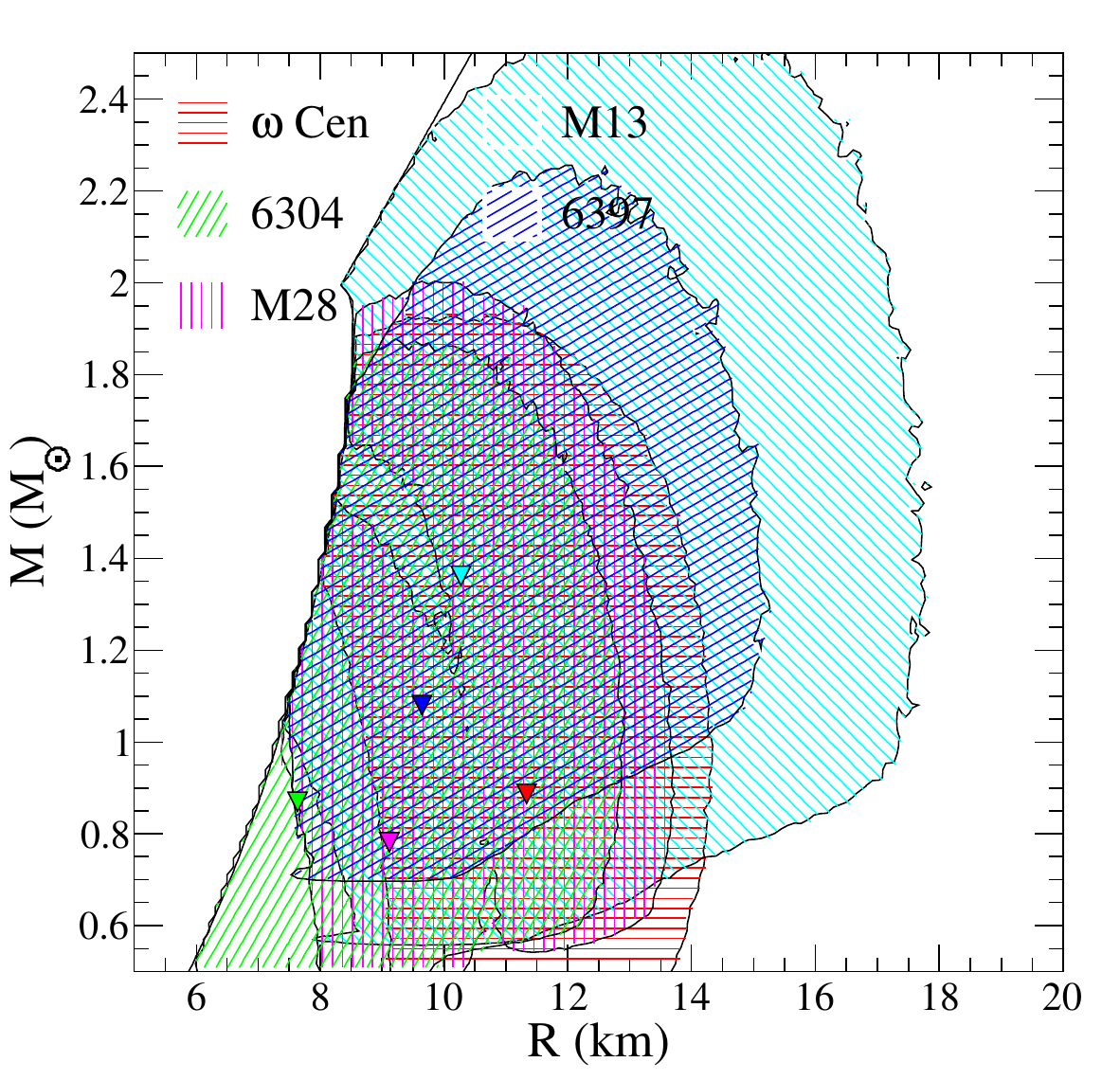}
\includegraphics[width=.49\textwidth]{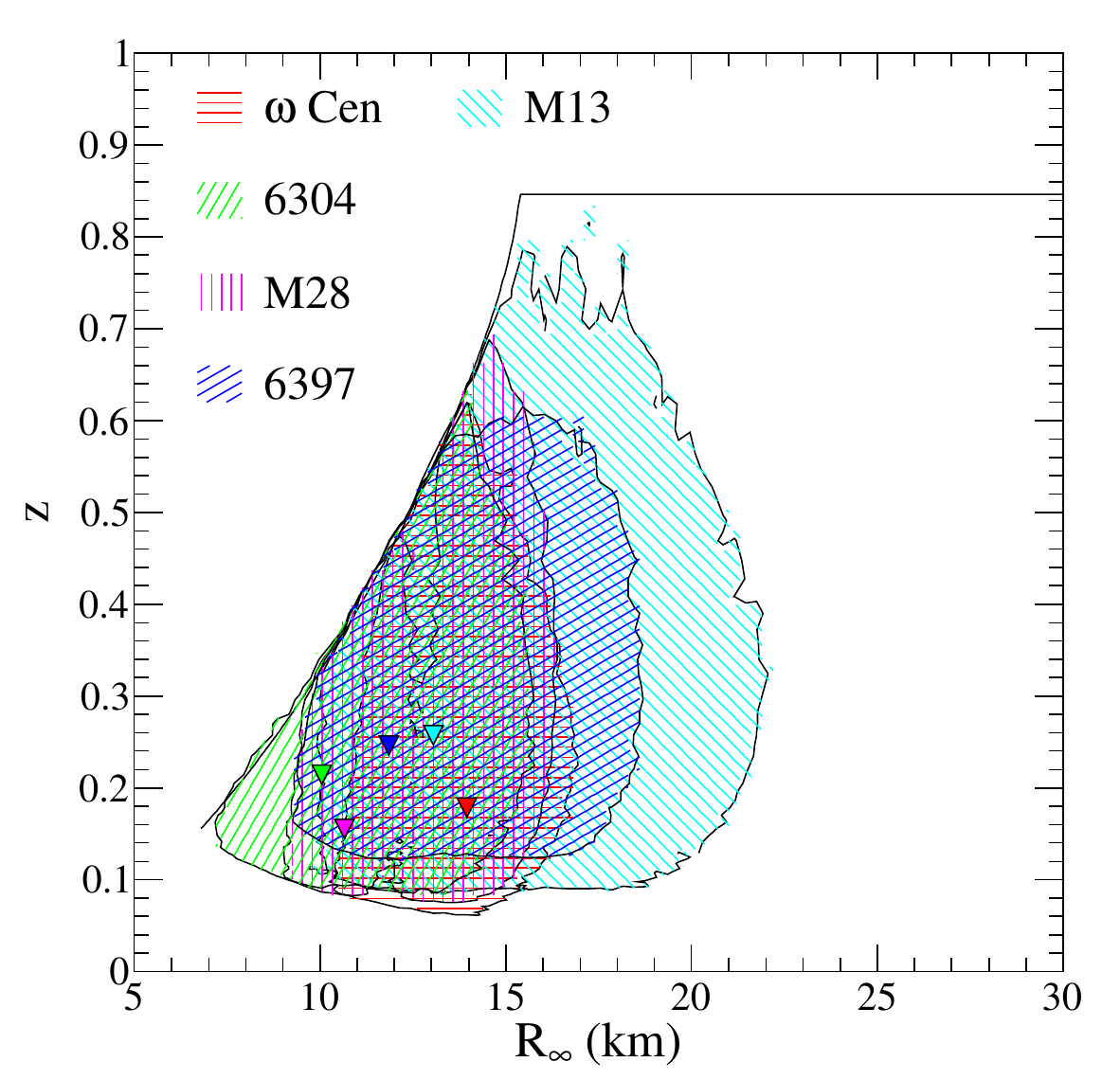}
\end{center}
\caption{Probability distributions in the $(z,R_{\infty})$ and $(M,R)$
  planes assuming the adjusted D90 $N_H$ values and G13 distances, but
  allowing the composition of the atmosphere to be either hydrogen or
  helium.
\label{fig:obs_He}}
\end{figure}

As discussed in \S \ref{Sec:helium}, we approximate the effect of a
helium atmosphere relative to a hydrogen atmosphere by scaling the
$M-R$ coordinates of the probability distributions of Figure
\ref{fig:corr} and the values of $R_{\infty}$ of column 6 of Table
\ref{tab:nh} by a further factor of 4/3 for all of the neutron stars
except that in $\omega$ Cen. We then add the He atmosphere probability
densities to those for the hydrogen atmosphere to get the full
distribution which allows for either composition. For the case of D90
$N_H$ values and G13 distances, the $M-R$ and $z-R_\infty$ probability
distributions are shown in Fig. \ref{fig:obs_He}. In the cases that
$R_{\infty}$ is sufficiently well determined by either assumed
composition, as in the case of M28, this gives a visibly bimodal form
to the $M-R$ distribution. We assume that hydrogen or helium
atmospheres are equally probable, so the integrals of the probability
density are equal. The helium region is broader because a larger range
for $R_{\infty}$ gives a larger mass range, so the peak of the helium
region must be lower.

\section{EQUATION OF STATE AND $M-R$ CURVES FROM QLMXB DATA}
\label{Sec:eosmr}

We now analyze the QLMXB data using the Bayesian method described in
\citet{Steiner10,Steiner13}. In this analysis, the
inferred $M-R$ probability distributions for these sources are
confronted with additional constraints imposed by the assumption that
the sources are neutron stars whose masses and radii are not free to
vary independently but are related through the relativistic equations
of stellar structure and contain a well-understood hadronic crust (the
stellar region with densities below about
$\rho_s/2\simeq1.3\cdot10^{14}$ g cm$^{-3}$). In addition, we impose
the constraints that the EOS is everywhere causal (i.e., that the
speed of sound cannot exceed the speed of light) and $\hat
M=2.0~M_\odot$. The EOS at densities higher than those of the crust is
divided into three regions. The baryon properties of the lowest of
these density regions is described by an expansions of the nucleon
energy per baryon in density and neutron excess around $\rho_s$ and
isospin symmetry. Electrons are treated as relativistic and
degenerate, with the composition (i.e., the electron and proton
fractions) determined by beta equilibrium. In the baseline model, the
uppermost two density regions are described with polytropes (the model
used in \citet{Steiner10} and also called model A in
\citet{Steiner13}) that define the pressure as a function of energy
density in beta equilibrium. This model is appropriate for neutron
stars without strong phase transitions and is labeled ``Base'' below.
For neutron stars with strong phase transitions due to the appearance
of exotic matter, such as from a quark-hadron phase transition, we
instead employ discrete line segments in the pressure--energy density
plane (model C in \citet{Steiner13} and labeled ``Exo'' below). The
assumed division of the high-density EOS into three regimes is amply
justified: \cite{Read09} has demonstrated that the detailed EOS of a
wide variety of strong interaction models is accurately predicted by
three polytropic segments. In either case (Base or Exo), the
theoretical uncertainties in the crustal EOS produce insignificant
changes to our conclusions. We also constrain neutron stars to be more
massive than $0.8~M_\odot$, a conservative lower limit. This lower
limit can be justified theoretically, but it should be noted there is
no significant observational evidence for the existence of neutron
stars with less than $1.1-1.2~M_\odot$.

In contrast, in their joint analysis of the five QLMXB sources, G13
did not allow the radii of the five individual neutron stars to freely
vary: a common value was determined by optimizing the spectral fits to
observational data. In their procedure, the correlations between
masses and radii resulting from the stellar structure equations were
not taken into account, and the strong constraints available from
knowledge of the low-density crust EOS and causality were not
considered. The lower limit to the neutron star mass was taken to be
$0.5~M_{\odot}$. G13 indirectly made use of the constraint that $\hat
M>2~M_{\odot}$ in that they justified their assumption of a near-constant
radius on the grounds that the discovery of $2~M_{\odot}$ neutron stars
favors interiors composed of "normal matter", rather than "quark
matter" or exotic matter with strong phase transitions \citep{LP10}.
Especially for EOSs with weak symmetry energy density dependence,
i.e., EOSs predicting that $1.4~M_{\odot}$ have radii smaller than
approximately 12 km, normal-matter stars with have less than a 10\%
range of radii for $M>0.5~M_{\odot}$ \citep{Lattimer01}. However, if the
EOS has a strong symmetry energy density dependence, such that
$1.4~M_{\odot}$ stars have $R\simge14$ km, the constant-radius assumption
becomes less valid.

In their joint analysis of the 5 QLMXB sources under the assumption
that all neutron stars have the same radius, G13 determined that the
most-probable value of this radius was $9.1^{+1.3}_{-1.5}$ km when
$N_H$ values were allowed to float and were simultaneously determined
from spectral fits. In the case that $N_H$ values were frozen at the
values optimized in the individual spectral fits, the joint analysis
yielded a common radius of $8.0\pm1.0$ km. In both cases, the sources
are determined to have optimum masses with a large variance, with
masses ranging from $0.72~M_{\odot}$ to $2.28~M_{\odot}$ when $N_H$ is
frozen, (Run 4 in G13) which would require either a large variation of
neutron star birth masses or a large amount of accretion. It is
interesting that these radii are relatively small compared to the
simple average (11.3 km) of the individual best-fit radii. Partly,
this is due to the fact that G13 weighted the contributions of
individual sources according to the quality of data from each source.
The source with the smallest inferred radius (NGC 6397) contained 35\%
of the total weight while the source with the largest inferred radius
($\omega$ Cen) contained 7.9\% of the total weight. With this unequal
weighting, the average radius is reduced to 9.8 km. This is, however,
still larger than the common radii G13 determined (although marginally
consistent to 90\% confidence with the result from their 'floating
$N_H$' analysis).

There is a straightforward explanation of this result. The values of
$N_H$ and $R_\infty$ determined for the 5 QLMXBs in either joint
analysis are not significantly different from their values determined
in the individual spectral fits. $R_\infty$ is more accurately
determined than $z$. If the values of $R_\infty$ are kept fixed for
each source, with the values determined in the individual spectral
fits, values of the orthogonal variable $z$ change to force a common
radius $R$. This suggests that one could estimate the common radius,
$R$, for the 5 QLMXBs by minimizing the function
\begin{equation}\label{one}
    \chi^2=\prod_i\left\{\exp\left(-w_i\left[{z(R,R_{\infty,i})-
        z(R_i,R_{\infty,i})\over\Delta z_i}\right]^2\right)\right\}
\end{equation}
with respect to $R$, where $w_i$ is the weight
associated with source $i$, and the values of the individual fits for
$R$, $R_\infty$ and the 90\% confidence interval for $z$ of the $i$th
source, are $R_i,~R_{\infty,i}$ and $\Delta z_i$, respectively. Given
the definition
\begin{equation}\label{two}
z(R,R_\infty)={R_\infty\over R}-1,
\end{equation}
minimization leads to
\begin{equation}\label{three}
\sum_i\left[{2w_iR_{\infty,i}\over R^2\Delta z_i}
\left({z(R,R_{\infty,i})-z(R_i,R_{\infty,i})\over 
\Delta z_i}\right)\right]=0,
\end{equation}
or, solving for $R$,
\begin{equation}\label{four}
R=\sum_i{w_iR_{\infty,i}^2\over\Delta z_i^2}\Bigg/
\sum_i{w_iR_{\infty,i}^2\over R_i\Delta z_i^2}.
\end{equation}
Note that normalizations associated with the Gaussian distributions in
Equation (\ref{one}) will cancel from Equation (\ref{three}).
Taking values for $R_i$ and $R_{\infty,i}$ from G13, and $\Delta z_i$
from Figure \ref{fig:uncorr}, we find the common radius to be
$R=8.1$ km. This value is in excellent agreement with G13's
value ($8.0\pm1.0$ km) when $N_H$ is frozen. 

An advantage of our Bayesian analysis is that it permits one to
compare in an unbiased way the quality of models with different prior
assumptions in terms of their fits to the observations. We can, for
example, compare models assuming normal-matter neutron stars as
opposed to quark matter stars. Or, we can compare models with
different assumptions about absorption, distance and composition. In
addition, our analysis leads to an explicit prediction for the $M-R$
curve, so that the validity of the constant-radius assumption can be
ascertained. In the context of Bayesian statistics, a commonly
accepted way of comparing two models is through the use of Bayes
factors. Given a particular model, ${\cal M}_{\alpha}$, we can define
the integral $I_{\alpha}$ (sometimes called the evidence) as
\begin{eqnarray}
I_{\alpha} &\equiv& \int P[{\cal D}|{\cal M_{\alpha}}] d
p^{(\alpha)}_{1}~dp^{(\alpha)}_{2} ... dp^{(\alpha)}_{N_p} \nonumber \\
&& \times d M_1~dM_2 ... dM_{5} 
\end{eqnarray}
where $P[{\cal D}|{\cal M_{\alpha}}]$ is the conditional probability
determined by the data and $N_p$ is the number of parameters (denoted
$p^{(\alpha)}$) in the EOS parameterization. This notation is defined
and further discussed in \citet{Steiner10}. The outer integrations are
over the neutron star masses (in this case, for five objects). (This
integral is simplified because we assume uniform priors in both the
EOS parameters and the neutron star masses.) The Bayes factor for
comparing two possible models ${\cal M}_{\alpha}$ and ${\cal
  M}_{\beta}$ is then $B_{\alpha,\beta} = I_{\alpha}/I_{\beta}$.
Typically, a Bayes factor of 3 would represent ``substantial''
evidence that model ${\cal M}_{\alpha}$ is preferred to model ${\cal
  M}_{\beta}$, a Bayes factor of 10 would be ``strong evidence'', and
100 would be ``decisive'' for model ${\cal M}_{\alpha}$ over ${\cal
  M}_{\beta}$. Values less than one give the opposite conclusion, e.g.
a Bayes factor less than 1/10 would be strong evidence for model
${\cal M}_{\beta}$ over ${\cal M}_{\alpha}$. 

\begin{deluxetable}{llllrr}
\tabletypesize{\scriptsize}
\tablecaption{Posterior Confidence Ranges and Evidence Integrals
\label{tab:bf}}
\tablehead{
\colhead{Model} & \colhead{$N_H$} & \colhead{Dist.} & \colhead{Comp.} &
\colhead{$R_{1.4}$ (km)} & \colhead{$I$}
}
\startdata
Base & G13 & G13 & H    & 11.11$-$11.88 & (1.77 $\pm$ 0.09) $\times 10^{-8}$ \\
Base & G13 & G13 & H+He & 11.36$-$12.84 & (4.50 $\pm$ 0.21) $\times 10^{-3}$ \\
Base & G13 & Alt & H    & 10.73$-$11.65 & (1.86 $\pm$ 0.18) $\times 10^{-6}$ \\
Base & G13 & Alt & H+He & 11.45$-$13.32 & (3.71 $\pm$ 0.21) $\times 10^{-1}$ \\
Base & G13 & H10 & H    & 10.77$-$11.71 & (1.23 $\pm$ 0.09) $\times 10^{-7}$ \\
Base & G13 & H10 & H+He & 11.36$-$13.44 & (4.28 $\pm$ 0.35) $\times 10^{-3}$ \\
Base & D90 & G13 & H    & 10.67$-$11.51 & (4.65 $\pm$ 0.48) $\times 10^{-3}$ \\
Base & D90 & G13 & H+He & 11.31$-$12.64 & (2.14 $\pm$ 0.19) $\times 10^{+2}$ \\
Base & D90 & Alt & H    & 10.85$-$11.79 & (9.40 $\pm$ 1.22) $\times 10^{-3}$ \\
Base & D90 & Alt & H+He & 11.37$-$12.61 & (4.06 $\pm$ 0.36) $\times 10^{+2}$ \\
Base & D90 & H10 & H    & 10.78$-$11.70 & (4.78 $\pm$ 0.73) $\times 10^{-3}$ \\
Base & D90 & H10 & H+He & 11.23$-$12.62 & (1.57 $\pm$ 0.07) $\times 10^{+2}$ \\
Base & H10 & G13 & H    & 10.87$-$11.82 & (1.04 $\pm$ 0.08) $\times 10^{+0}$ \\
Base & H10 & G13 & H+He & 11.15$-$12.38 & (1.84 $\pm$ 0.12) $\times 10^{+2}$ \\
Base & H10 & Alt & H    & 11.03$-$12.07 & (1.39 $\pm$ 0.20) $\times 10^{+2}$ \\
Base & H10 & Alt & H+He & 11.04$-$12.31 & (1.44 $\pm$ 0.10) $\times 10^{+2}$ \\
Base & H10 & H10 & H    & 10.78$-$11.95 & (7.52 $\pm$ 0.65) $\times 10^{+1}$ \\
Base & H10 & H10 & H+He & 11.31$-$12.66 & (5.30 $\pm$ 0.22) $\times 10^{+2}$ \\
Exo & G13 & G13 & H     &  9.15$-$10.81 & (7.32 $\pm$ 0.63) $\times 10^{-6}$ \\
Exo & G13 & G13 & H+He  & 10.52$-$11.77 & (4.46 $\pm$ 0.38) $\times 10^{-2}$ \\
Exo & G13 & Alt & H     & 10.42$-$11.39 & (1.21 $\pm$ 0.19) $\times 10^{-3}$ \\
Exo & G13 & Alt & H+He  & 10.88$-$12.59 & (7.33 $\pm$ 0.78) $\times 10^{-1}$ \\
Exo & G13 & H10 & H     & 10.61$-$11.41 & (2.23 $\pm$ 0.48) $\times 10^{-5}$ \\
Exo & G13 & H10 & H+He  & 10.76$-$12.38 & (1.67 $\pm$ 0.16) $\times 10^{-2}$ \\
Exo & D90 & G13 & H     &  9.39$-$10.97 & (5.46 $\pm$ 1.74) $\times 10^{-1}$ \\
Exo & D90 & G13 & H+He  & 10.53$-$12.45 & (2.29 $\pm$ 0.13) $\times 10^{+1}$ \\
Exo & D90 & Alt & H     &  9.86$-$11.44 & (3.04 $\pm$ 0.42) $\times 10^{-1}$ \\
Exo & D90 & Alt & H+He  & 10.90$-$12.31 & (4.46 $\pm$ 0.22) $\times 10^{+1}$ \\
Exo & D90 & H10 & H     &  9.60$-$11.38 & (2.27 $\pm$ 0.50) $\times 10^{-1}$ \\
Exo & D90 & H10 & H+He  & 10.61$-$12.28 & (2.59 $\pm$ 0.15) $\times 10^{+1}$ \\
Exo & H10 & G13 & H     &  9.87$-$11.49 & (5.15 $\pm$ 0.51) $\times 10^{+0}$ \\
Exo & H10 & G13 & H+He  & 10.60$-$11.99 & (4.67 $\pm$ 0.46) $\times 10^{+1}$ \\
Exo & H10 & Alt & H     & 10.45$-$11.74 & (5.17 $\pm$ 0.64) $\times 10^{+1}$ \\
Exo & H10 & Alt & H+He  & 10.53$-$11.81 & (7.49 $\pm$ 0.75) $\times 10^{+1}$ \\
Exo & H10 & H10 & H     & 10.42$-$11.72 & (2.83 $\pm$ 0.21) $\times 10^{+1}$ \\
Exo & H10 & H10 & H+He  & 10.74$-$12.39 & (8.93 $\pm$ 0.47) $\times 10^{+1}$ \\
\enddata
\tablecomments{
The first four columns give the model designations, the next column
gives the posterior 90\% confidence limits for the radius of a 1.4
solar mass neutron star, and the last column gives the ``evidence'',
the integral necessary to compute the Bayes factor.}
\end{deluxetable}

\subsection{The  Simulations}
\label{Sec:baseline}

The chief results of our simulations with different prior assumptions
regarding column densities, distances, atmosphere composition, and the
high-density EOS are given in Table \ref{tab:bf} in terms of the
predicted radii of $1.4~M_{\odot}$ stars and the evidence integral I for
comparing models. Uncertainties in $I$ are due to the interpolation
used in the integration. The three choices of column density sets are
those from G13, D90, and H10. The three choices of distance sets, as
described in \S \ref{Sec:Data}, are those from G13, the "Alt" set, and
H10. The two assumed atmosphere compositions are H and the possibility
of either H or He. Finally, the two assumptions for the high-density
EOS are that they are given by polytropes, which represent matter
without a strong phase transition, labeled "Base", or by four line segments, which represent matter with strong
phase transitions typical of EOSs which contain exotic matter. Thus,
there are a total of 36 simulations.
\begin{figure}
\begin{center}
\includegraphics[width=.49\textwidth]{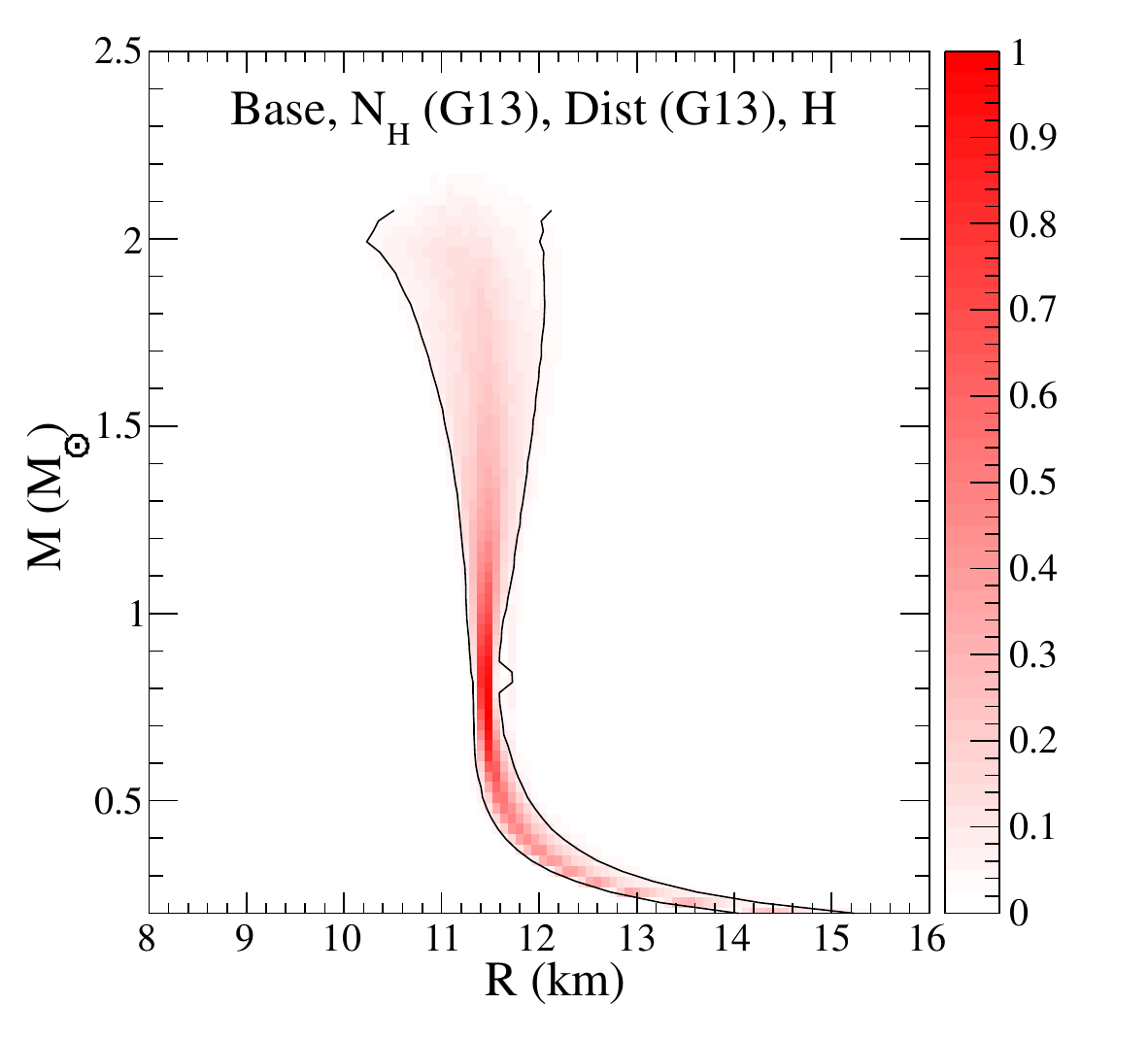}
\includegraphics[width=.49\textwidth]{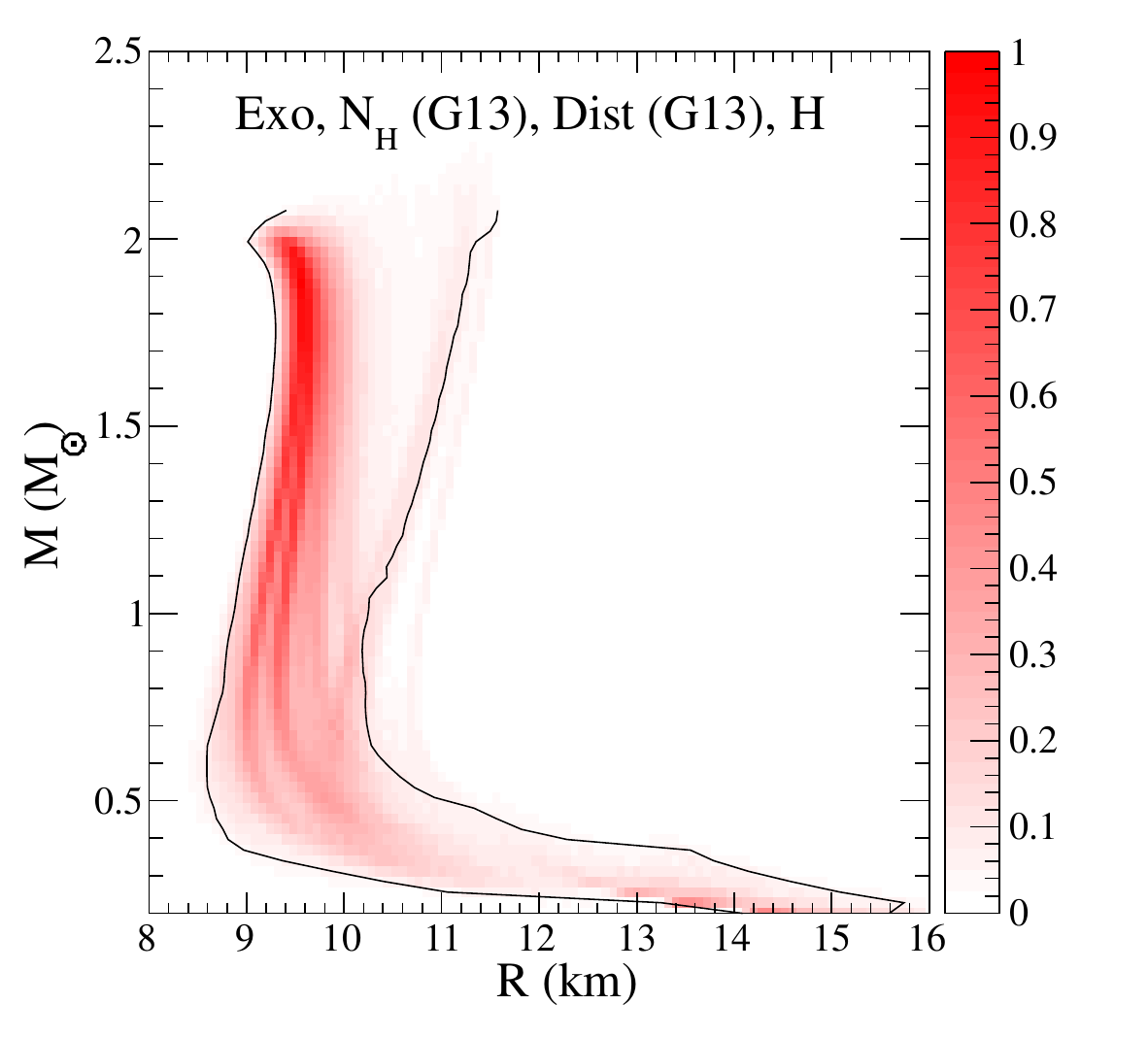}
\end{center}
\caption{The probability distributions for mass as a function of
  radius for neutron stars assuming the baseline EOS (left panel) or a
  model which favors strong phase transitions (right panel) based on
  the QLMXB data from G13.}
\label{fig:uncorr_mr}
\end{figure}

The first row in Table~\ref{tab:bf} assumes distances, column
densities, and the H atmosphere composition following the assumptions
of G13, slightly modified by folding in the distance uncertainty and
also removing probability regions excluded by causality and the
condition $\hat M\ge2.0~M_\odot$ (Figure \ref{fig:uncorr}). The
high-density EOS is assumed to be described by two polytropes. The
radius probability distribution as a function of mass found by our
Bayesian analysis is displayed in the left panel of
Fig.~\ref{fig:uncorr_mr}. The range of radii for $1.4~M_{\odot}$ stars
is also shown in Table \ref{tab:bf} and is $10.11-11.88$ km for this
case. This radius range is largely outside the 90\% confidence range
$7.6{\rm~km}<R<10.4{\rm~km}$ determined by G13 under the assumptions
that (i) the radii of all sources are equal and (ii) $N_H$ values are
allowed to float. This range is also completely outside the 90\%
confidence range $7.0{\rm~km}<R<9.0{\rm~km}$ found by G13 when $N_H$
values are frozen. We attribute this, in part, to the implicit use of
the stellar structure equations and incorporation of a crustal EOS in
our simulation and partly due to the weighting G13 assigned for each
source.

\begin{deluxetable}{ll}
\tablecaption{Some relevant Bayes factors\label{tab:bf2}}
\tablehead{
\colhead{Model A/Model B} & \colhead{Bayes factor of A in favor of B} \\
}
\startdata
Base/Exo              & \phm{(}4.73 $\pm$ 0.20 \\
$N_H$(H10)/$N_H$(D90) & \phm{(}1.57 $\pm$ 0.09 \\
$N_H$(H10)/$N_H$(G13) & (1.17 $\pm$ 0.09) $\times 10^{3}$ \\
D(Alt)/D(G13)         & \phm{(}1.82 $\pm$ 0.13 \\
D(H10)/D(Alt)         & \phm{(}1.05 $\pm$ 0.06 \\
H+He/H                & \phm{(}6.44 $\pm$ 0.49 \\
\enddata
\tablecomments{The Bayes factors comparing various scenarios, computed
  by forming ratios of sums of the relevant rows from
  Table~\ref{tab:bf}. Jeffrey's scale for the Bayes factor suggests
  that values greater than 3 represent ``substantial'' evidence,
  values greater than 10 represent ``strong'' evidence, and values
  greater than 100 represent ``decisive'' evidence for the model in
  the numerator as compared to the model in the denominator.}
\end{deluxetable}

However, the set of assumptions used in the first line of Table
\ref{tab:bf} represents a model with a relatively poor fit to the data
(the evidence integral $I$ for this case is 10 orders of magnitude
smaller than for the most-favored cases). This is caused by the fact
that the $M-R$ distributions found in the joint analysis for the
sources in $\omega$ Cen and NGC 6397 are well outside their most
probable regions when individually analyzed (compare our Figure
\ref{fig:uncorr} or Figures 4 -- 8 in G13 with Figures 9 -- 16 in
G13). For example, the most probable mass for the neutron star in
$\omega$ Cen is driven in the joint analysis to 2~\Msun and the radius
to 10 km, just at the edge of the region allowed by causality,
compared to 1.6~\Msun and 20 km, respectively, in its individual
analysis.

Our results for this case but assuming an exotic matter EOS (right
panel of Fig.~\ref{fig:uncorr_mr}, model ``Exo'') imply smaller radii,
$9.15-10.91$ km for a $1.4~M_{\odot}$ star, because of the presence of
strong phase transitions (as observed in \citet{Steiner13}). The
evidence integral, $I\approx 7 \times 10^{-6}$, is more than two
orders of magnitude larger than the model without strong phase
transitions, $I \approx 2 \times 10^{-8}$. This suggests that if the
G13 model for X-ray absorption and distance is correct, dense matter
is likely to exhibit some sort of strong phase transition. In order to
make a less model-dependent statement about phase transitions,
however, we should look at the ratio of the sums of the evidence
integrals for all the 18 cases denoted ``Base'' in Table~\ref{tab:bf}
and all evidence integrals for all 18 cases denoted ``Exo''. This
ratio is reported in the top row of Table~\ref{tab:bf2} as 4.7. Thus,
over all combinations of assumptions of distances, $N_H$ values, and
atmosphere compositions, models without strong phase transitions are
moderately favored. This is opposite to the above conclusion obtained
from examining only one model, highlighting the importance of
considering several models before making a definitive conclusion.

\subsection{Alternate $N_H$ and Alternate Distance Simulations}
\label{Sec:altnh}

\begin{figure}
\begin{center}
\includegraphics[width=.49\textwidth]{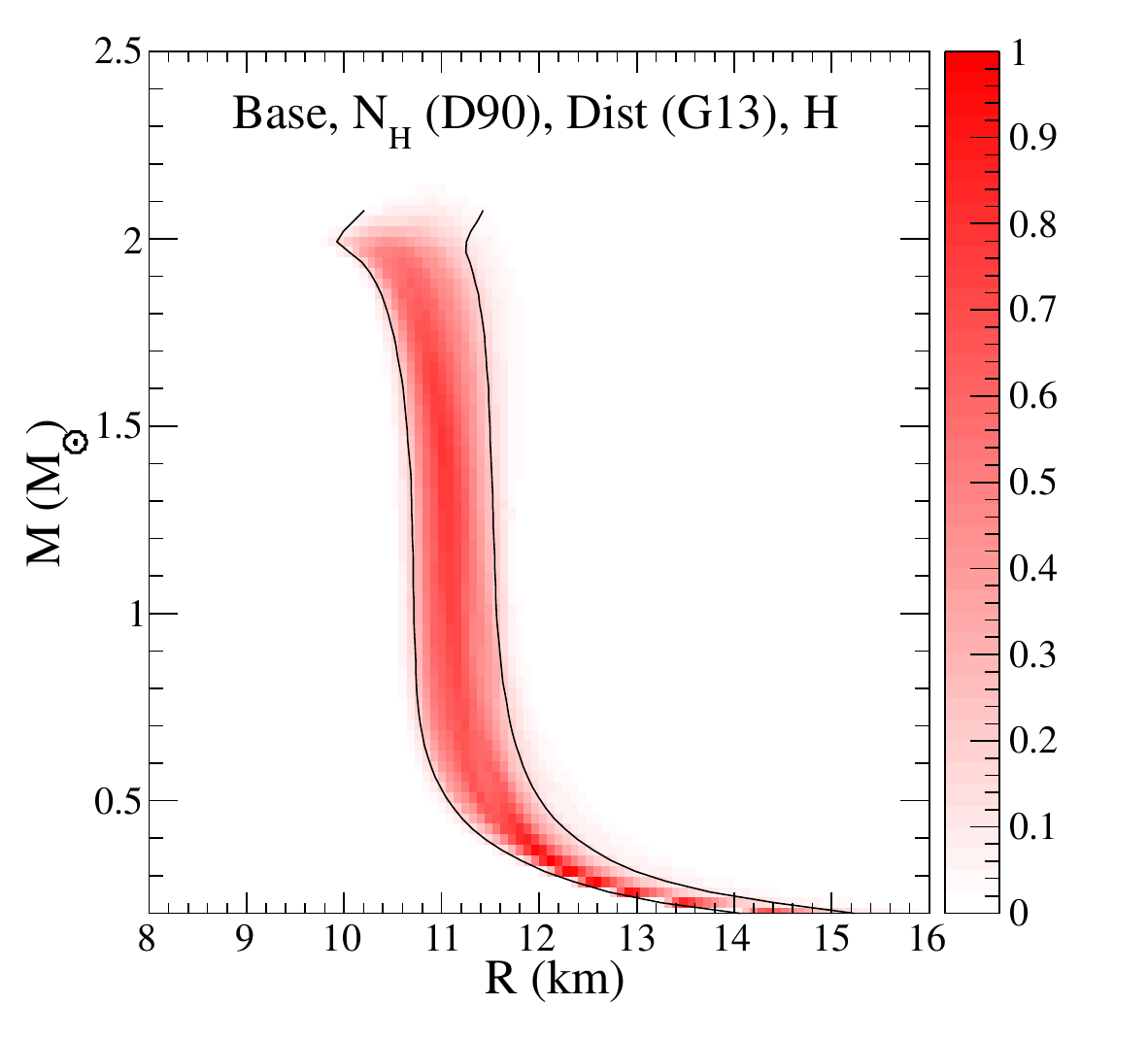}
\includegraphics[width=.49\textwidth]{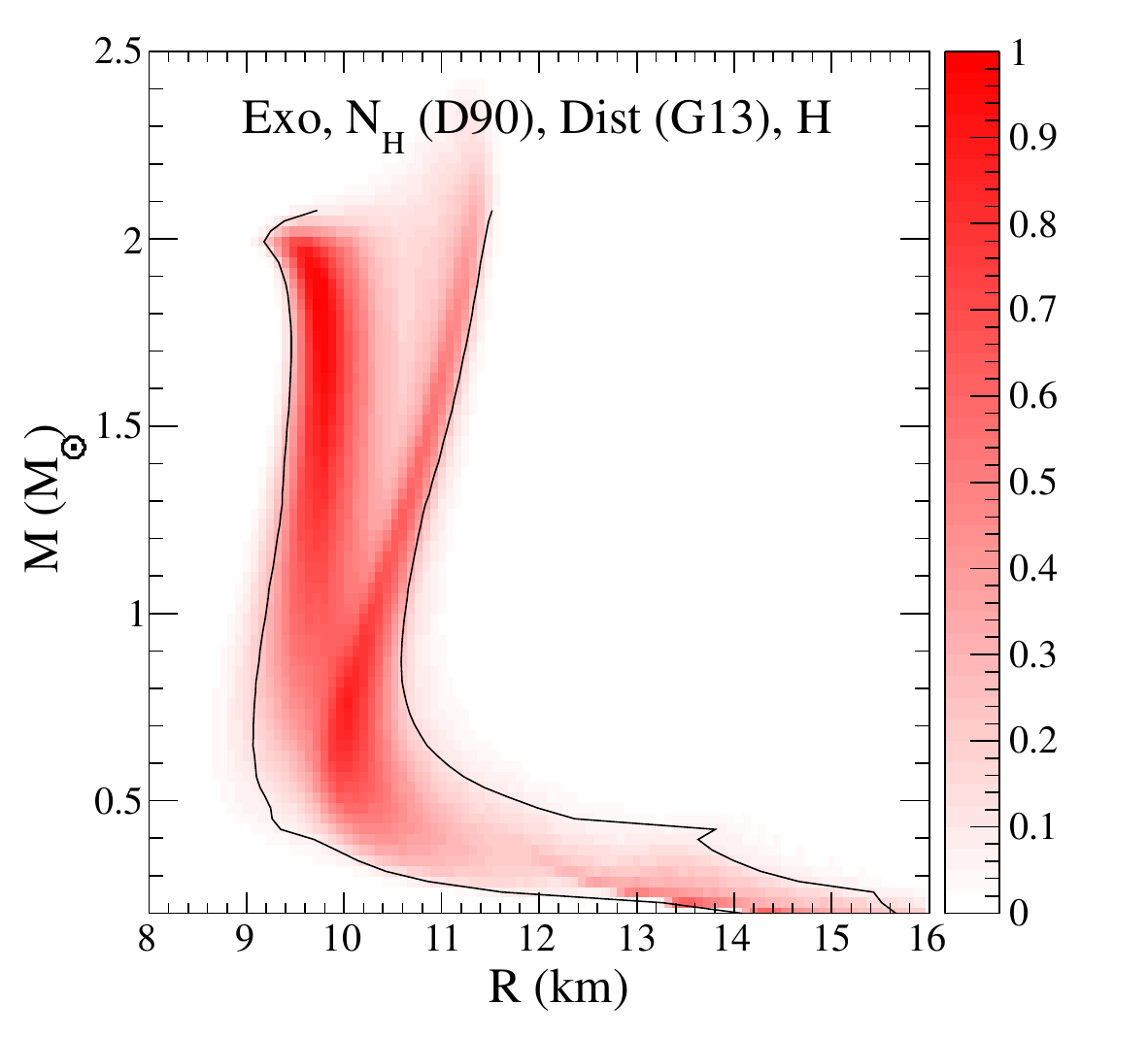}
\end{center}
\caption{The probability distributions for mass as a function of
radius for neutron stars for the baseline EOS (left panel) and 
for the EOS with strong phase transitions
based on the QLMXB data from G13 after the adjustment
 using the $N_H$ values from D90.}
\label{fig:alt_mr}
\end{figure}

We now consider models with the alternative assumptions
concerning $N_H$. For the case of the D90 $N_H$
values, the corresponding $M-R$ curves, after having modified
$R_{\infty}$ according to the prescription described in \S
\ref{Sec:altint}, are displayed in
Fig.~\ref{fig:alt_mr}. We again find that the assumption of an
EOS with strong phase transitions leads to smaller predicted radii than
for the Base EOS.  The average difference in radii is about 0.5 km, but
the 90\% confidence range is more than doubled in the Exo case.

After allowing for any of the distance sets, EOS models, and
atmosphere compositions (with the exception of $\omega$ Cen), the
Bayes factor for H10 in favor of D90 is 1.57 (Table \ref{tab:bf2}),
showing no strong preference between the two models. However, the
Bayes factor for H10 in favor of G13 hydrogen column densities is over
1000. This demonstrates that, unless there is some other important
model uncertainty which we have not considered, the set of X-ray
absorptions determined by G13 appears ruled out. We also find little
evidence to suggest that any of the three distance sets is preferred
as the associated Bayes factors are all of order unity, as shown in
fourth and fifth rows of Table~\ref{tab:bf2}.

\begin{figure}[h]
\begin{center}
\includegraphics[width=.49\textwidth]{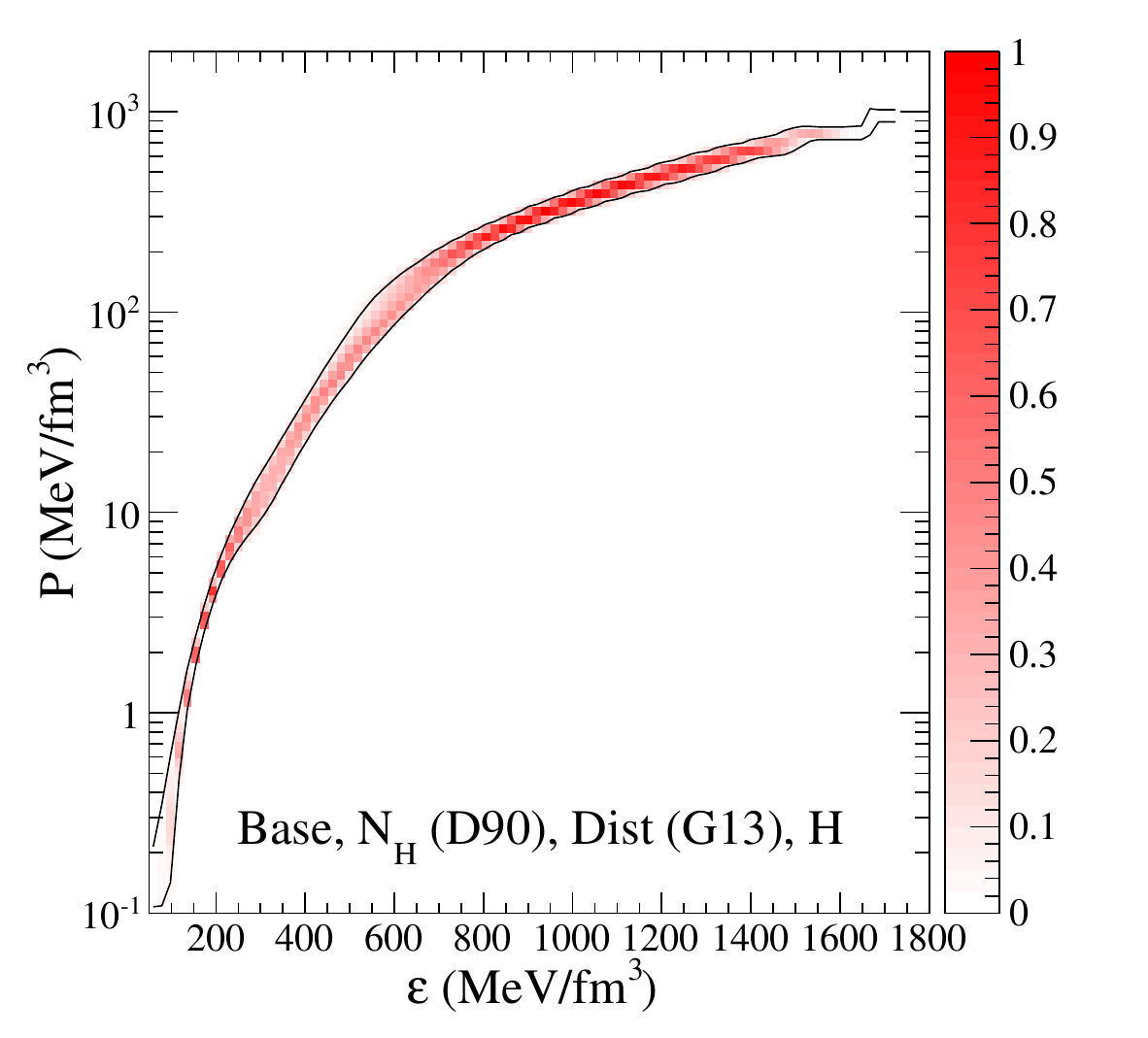}
\includegraphics[width=.49\textwidth]{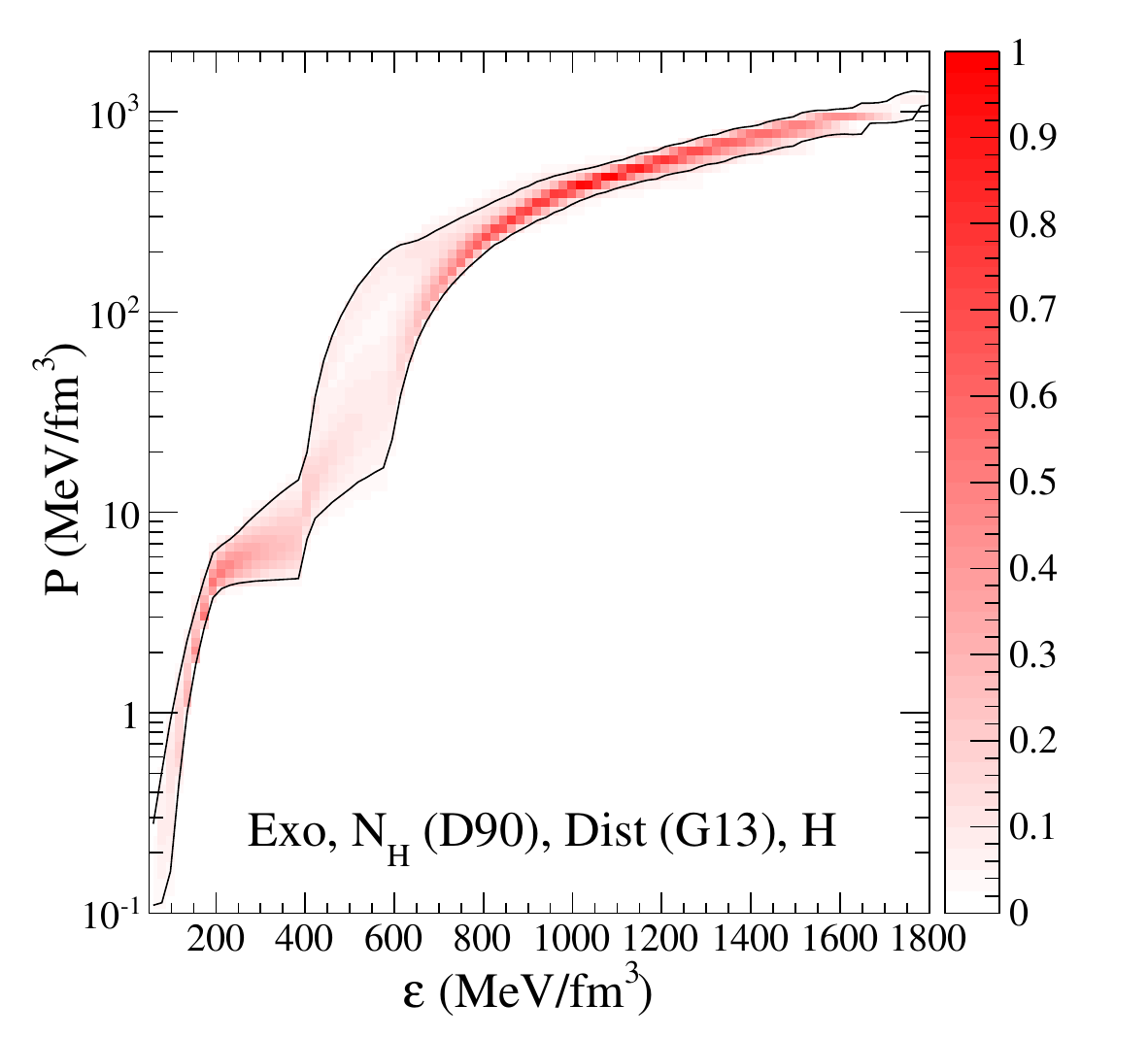}
\end{center}
\caption{The probability distributions for the pressure as a function
  of energy density corresponding to the QLMXB data from G13 after the
  $R_{\infty}$ correction due to the alternate $N_H$ values. The left
  panel displays results for the Base EOS and the
  right for the Exo EOS. }
\label{fig:corr_eos}
\end{figure}

The Bayesian analysis not only leads to a predicted $M-R$ curve for
each set of assumptions concerning the underlying EOS, distance and
absorption, but also allows the prediction of the EOS parameter and
the resulting pressure--energy density relation. Since our preferred
Base and Exo models always prefer alternate values of $N_H$, we
compare such Base and Exo predictions for the EOS in
Fig.~\ref{fig:corr_eos}. Both figures show the 90\% confidence limits.
The kinks in the pressure-density relation for the Exo model are the
result of phase transitions which allow the radius to be smaller for
low-mass neutron stars.

\subsection{Simulations with both H and He Atmospheres}
\label{Sec:eosHe}

\begin{figure}[h]
\begin{center}
\vspace*{-1cm}
\includegraphics[width=.49\textwidth]{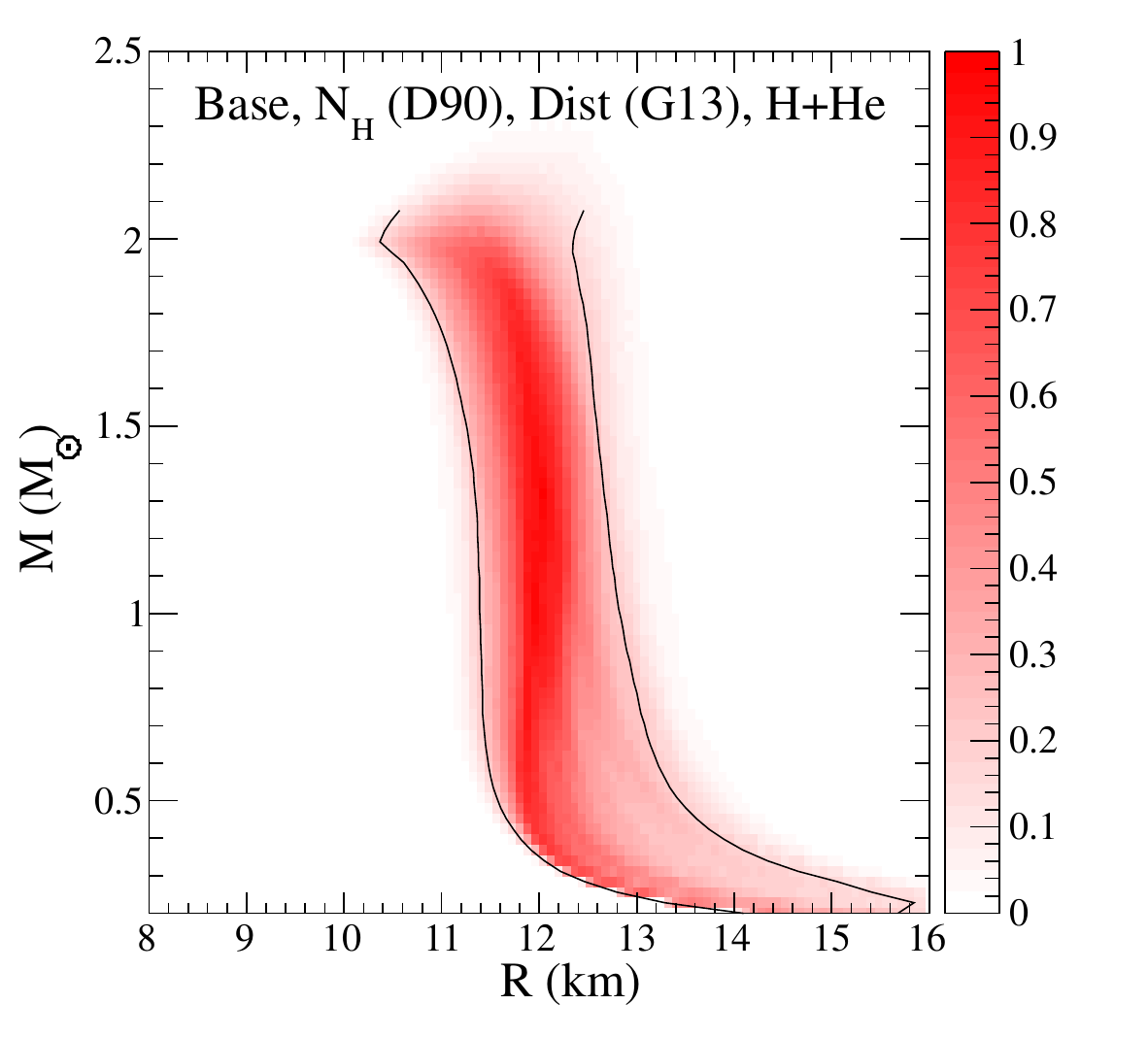}
\includegraphics[width=.49\textwidth]{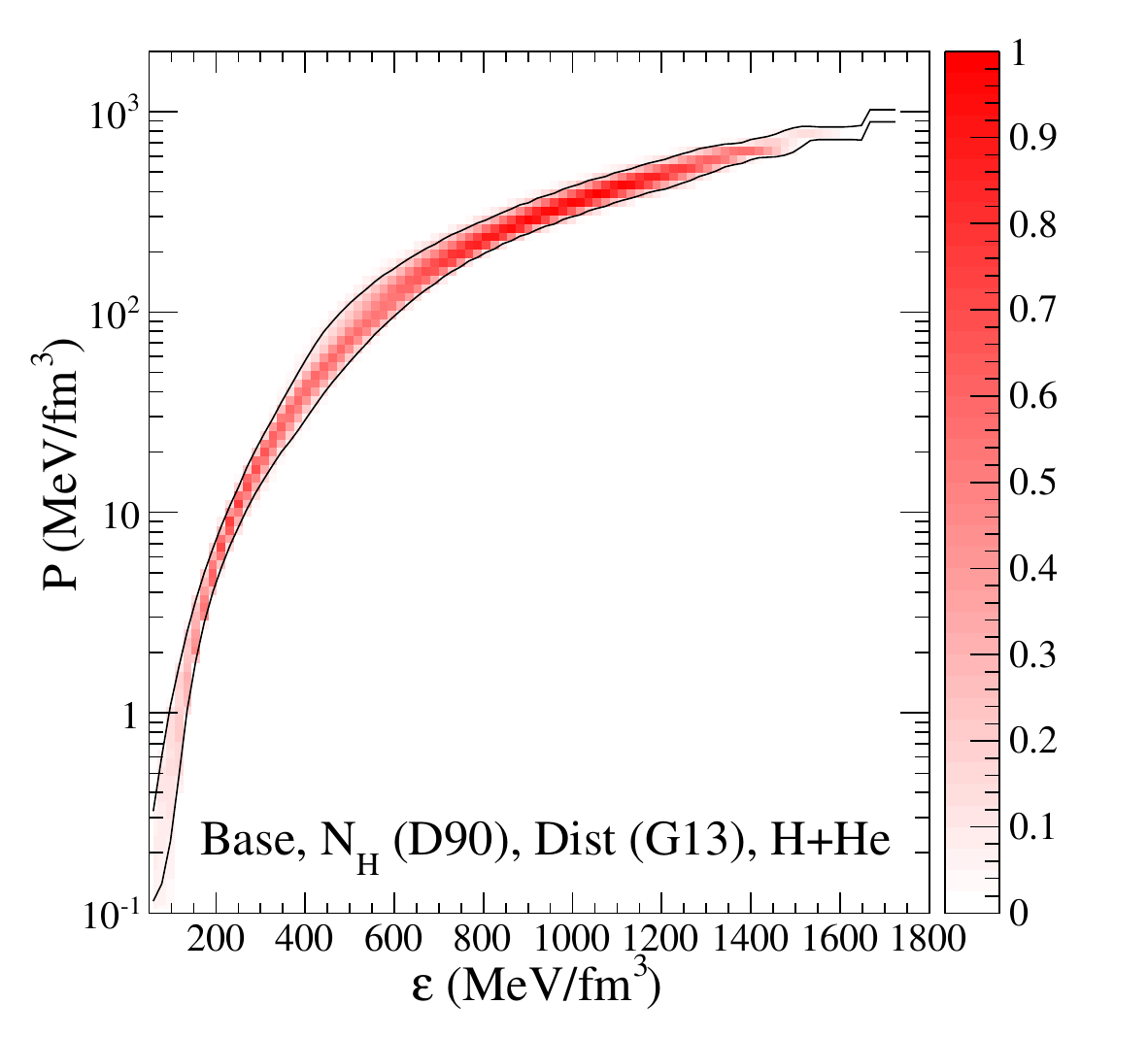}
\end{center}
\caption{Mass-radius (left panel) and EOS probability distributions
  (right panel) for neutron stars assuming the baseline
  EOS, with alternate set of column densities and allowing for
  hydrogen or helium atmospheres.}
\label{fig:He_mreos}
\end{figure}

Our results indicate there is evidence that at least one of the
neutron stars other than $\omega$ Cen has a helium atmosphere. This
assertion is supported by the overall Bayes factor for the H+He in
favor of the H models, which is 6.4, as given in the last row of
Table~\ref{tab:bf2}. In the particular case where there are no strong
phase transitions, the $N_H$ values are given by D90, and the
distances are taken from G13, the Bayes factor is considerably larger,
$5\times10^4$. The associated EOS and $M-R$ curves are given in
Fig.~\ref{fig:He_mreos}, and the radius of a 1.4~\Msun neutron star is
between 11.31 km and 12.64 km. For the H10 values of $N_H$, the Bayes
factor is about 180 and the predicted radius range is about 0.3 km
lower. In either case, these radii are similar to those that one
expects from the PRE X-ray sources~\citep{Steiner13}. However, these
conclusions depend on assumptions about $N_H$. If a future study were
to confirm even larger values of $N_H$ for M28 and NGC 6304 than those
of H10, then the evidence for helium atmospheres would become weaker.

\section{DISCUSSION AND CONCLUSIONS}        
\label{Sec:discussion}

Our results show that models employing independently determined values
for $N_H$ are strongly favored over models that use $N_H$ values
self-consistently derived as part of the spectral fitting procedure.
We also find that allowing the possibility of either hydrogen or
helium atmospheres is strongly favored by the currently available mass
and radius data from QLMXBs. In this case, the combination of
independently-determined $N_H$ values with the possibility of either H
or He atmospheres, there is substantial evidence that the
EOS of dense matter does not have a strong phase transition, such as
those due to the quark-hadron phase transition. However, if all
sources are eventually shown to possess hydrogen atmospheres, the EOS
is then favored to have strong phase transitions. For our Base EOS and
over all assumptions about distance, absorption, and atmosphere
composition, the models which have the largest value of the
evidence integral, $I > 10^{2}$, suggest that the radius of a
$1.4~M_{\odot}$ neutron star is predicted to be between 11.15 and
12.66 km. For the Exo EOS, the models which have the
largest evidence, $I > 10^{1}$, give smaller radii, between 10.45 and
12.45 km. Both of these ranges are consistent with the predicted
radius range for $1.4~M_\odot$ stars from nuclear experimental
\citep{Newton11,Tsang12,Lattimer13} and theoretical neutron matter
studies \citep{Steiner12,Hebeler13}, about 10.7--13.1 km and 9.7 --
13.9 km, respectively, with all confidence regions being 90\%.

However, the uncertainties in $N_H$ might be quite large. The analysis
from G13 suggests that the ratio of $N_H$ values determined from
spectral fitting to those from HI surveys can range from 1/2 to 2. In
the context of mass and radius observations, this uncertainty has
enormous implications. Our work should motivate more extensive
observations in several wavelength regimes to determine $N_H$ along
the lines of sights to globular clusters with more precision. In the
radio, more detailed measurements of HI column densities and
metallicities of intervening matter are possible with present
technologies. In X-rays, however, the problem is more challenging
because large throughput detectors with high spectral resolution at
low energies (0.1 - 0.3 keV) are required to determine column
densities directly through observations of edges in the spectrum. It
is possible that observations of bright X-ray bursts from these
clusters could provide the required information. The absolute flux
calibration of X-ray detectors could also shift radius measurements in
either direction by as much as 15\%. Distance uncertainties are still
as large as 25\% and should be improved. Better X-ray data is needed
to determine the atmosphere compositions of accreting neutron stars in
QLMXB systems, as this can make 30\% or greater changes in inferred
neutron star radii. In addition observations of H$\alpha$ emission
could help pin down atmosphere compositions. The relative
normalizations of the QLMXB $M-R$ distributions could be varied and
such variation will change the Bayes factors and the associated
interpretation. These normalizations will be additionally confounded
by systematics which are common to all sources, such as those which
result from the slow decay of X-ray observing instruments.

There are also systematic uncertainties in the probability
distributions for neutron star radii, the EOS, and for the Bayes
factors coming from the choice of the prior distribution. The choice
of prior manifests itself in two ways: the selection of the EOS
parameterization, and the choice of the neutron star mass function.
The effect of the EOS parameterization could be analyzed more
systematically, e.g. through a hierarchical analysis, but this would
require a large computational effort beyond the scope of this work.
The neutron star mass function could also be varied, in line with
recent progress in mass measurements~\citep{Lattimer12} and this will
be pursued in future work. Assuming that neutron stars of low mass are
more probable will tend to prefer smaller values of $R_{\infty}$,
because causality and the maximum mass constraint tend to prefer $M-R$
curves which are vertical (i.e. fixed $R$) in the region of interest.
This could provide evidence in favor of strong phase transitions and H
atmospheres.

The thesis that low-mass X-ray binary systems studied here contain
strange quark stars rather than neutron stars could be consistent with
the relatively small range of redshifts and small radii obtained for
some of the QLMXBs especially if our alternate $N_H$ values are
correct for $\omega$ Cen. On the other hand, it is difficult for
strange quark stars to reproduce the wide array of phenomenology
observed in LMXBs including X-ray bursts, superbursts~\citep{Page05},
and crust cooling~\citep{Stejner06}. Furthermore, if the neutron star
maximum mass is substantially higher than $2~M_{\odot}$, as perhaps
indicated by the two black widow pulsar systems PSR B1957+20
\citep{vankerkwijk11} and PSR J1311-3430 \citep{Romani12}, as well as
the binary pulsar J1748-2021B in NGC 6440 \citep{Freire08}, the
possibility of strange quark stars is strongly disfavored.

This work is supported by DOE grants DE-AC02-87ER40317 (J.M.L.) and
DE-FG02-00ER41132 (A.W.S.). This work used the resources of the
National Energy Research Scientific Computing Center (NERSC), which is
supported by the Office of Science of the U.S. Department of Energy
under Contract No. DE-AC02-05CH11231. We thank E. Brown, S. Gandolfi,
S. Guillot and S. Reddy for discussions, S. Gandolfi for help with the
NERSC computations, S. Guillot for providing data from G13, and E.
Brown bringing to our attention the analytic approximation for
hydrogen atmospheres, Eq. (\ref{planck1}), and its derivation in Appendix A. 

\section*{APPENDIX A: Justification for the Value $p=5/7$\label{APP:a}}
To justify that $p=5/7$, consider the competition between scattering
and absorption in a hydrogen atmosphere. The electron scattering cross
section $\sigma_{s}$ is constant and, for energies near the peak of
the spectrum, is much greater than the free-free cross section
$\sigma_f$, which depends on energy, density and temperature as
$\sigma_f\propto\rho T_{\mathrm{eff}}^{-1/2}E^{-3}$. The total distance a photon
travels before being absorbed is approximately
$\lambda_f=N\lambda_{s}$ where $\lambda_{s,f}=(n_e\sigma_{s,f})^{-1}$
is the respective mean free path and $n_e$ is the number density of
electrons. For a random walk, the physical depth a photon travels
before being absorbed is
$z=\lambda_s\sqrt{N}=\sqrt{\lambda_f\lambda_s}$.

To allow for changes in density and temperature in the atmosphere,
emerging photons of energy $E$ originate from an approximate depth
determined by
\be\label{depth}
\tau(E)={N_A\over\mu_e}\int\sqrt{\sigma_s\sigma_f}\rho~dz\simeq1,
\ee
where $N_A$ is Avogadro's number and $\mu_e$ is the mean molecular
weight per electron. From hydrostatic equilibrium, $\rho~dz=dP/g$
where $g$ is the constant surface gravity and $P$ is the pressure. For
a gray atmosphere, $P\propto T_{\mathrm{eff}}^4\propto\rho T_{\mathrm{eff}}$. Eq. (\ref{depth}) can
thus be expressed as
\be
\label{depth1}
\int E^{-3/2}T_{\mathrm{eff}}^{17/4}dT_{\mathrm{eff}}\propto1.
\ee
Therefore, the temperature $T_a$ at the depth where emergent photons
of energy $E$ originate scales with energy as $T_a\sim E^{2/7}$. For a
Planckian spectrum at large optical depths, the specific flux therefore
behaves like
\be\label{intensity}
F_E\propto E^3\left[e^{E/kT_a}-1\right]^{-1}\propto E^3
\left[e^{\beta(E/kT_{\mathrm{eff}})^{5/7}}-1\right]^{-1}.
\ee
Calibrating the peak flux to model hydrogen atmospheres
\citep{Romani87,Zavlin96} allows determination of the specific flux in
this approximation,
\begin{equation}\label{flux}
F_E=8.73\cdot10^{22}~T_{\mathrm{eff}}^{0.2}E^3\left[e^{\beta(E/T_{\mathrm{eff}})^{5/7}}-1\right]^{-1}
{\rm~erg~cm}^{-3}{\rm~s}^{-1}{\rm~keV}^{-1},
\end{equation}
where $E$ and $T_{\mathrm{eff}}$ are in keV. Another analytic approximation was
determined by \cite{McClintock04}, which is
\begin{equation}\label{flux1}
F_E=5.26\cdot10^{23}~T_{\mathrm{eff}}^{0.5}E^{2.5}e^{-\beta^\prime(E/T_{\mathrm{eff}})^{0.55}}
{\rm~erg~cm}^{-3}{\rm~s}^{-1}{\rm~keV}^{-1},
\end{equation}
where $\beta^\prime=3.573$. Our approximation represents a better fit
for $T_{\mathrm{eff}}>10^6$ K, which suggests an improved
approximation might be found if $p$ was a monotonically increasing
function of temperature. However, most of the sources under study have
$T_{\mathrm{eff}}>10^6$ K, so we forgo a better approximation and
simply utilize Eq. (\ref{flux}) in the subsequent discussion. A
simplification afforded by either of Eqs. (\ref{flux}) or
(\ref{flux1}) is that effects of gravity will be straightforward to
approximate, which ceases to be the case when $p$ is a
temperature-dependent parameter.

\section*{APPENDIX B: Rescaling $R_{\infty}$ for the Neutron Star in 
$\omega$ Cen\label{APP:b}}

G13 obtained a large value for $R_\infty$ in the case of $\omega$ Cen
because they deduced a large value for $N_H$. Their probability
distributions were confined to $M<3~M_{\odot}$, which effectively
decreases to $M<1.8~M_{\odot}$ when $R_{\infty}$ is corrected by the
factor of 0.511 for our lower alternative value of $N_H$ as given in
Table 2. This results in missing information which creates an
unphysical constraint on the mass of the neutron star when computing
the corrected $R_{\infty}$ distribution. We therefore simulate data
for $M>1.8~M_{\odot}$ in this case using the same distribution in $z$ and
$R_{\infty}$ as that inferred from G13 in the region
$R_{\infty}<23.02$ km, the region unaffected by the $M<3~M_{\odot}$
limit. The simulated data is added to the original G13 data and the
sum is renormalized to ensure that the probability distribution is
smooth across the $M=1.8$~\Msun boundary.


\end{document}